\documentclass[letterpaper,11pt]{article}
\usepackage[utf8]{inputenc}

\pdfoutput=1

\usepackage{amsmath, amsthm, amssymb}
\usepackage[linesnumbered,vlined]{algorithm2e}
\usepackage[table,xcdraw]{xcolor}
\usepackage{mathtools}
\usepackage[numbers]{natbib}
\usepackage{comment} 
\usepackage{color-edits} 
\usepackage[most]{tcolorbox}
\usepackage{xfrac}
\usepackage{hyperref}
\usepackage{multirow}
\usepackage{caption}
\usepackage{bm}
\usepackage{newfloat}
\usepackage{enumitem}
\usepackage{xpatch}
\usepackage{soul}
\usepackage{bbm}

\makeatletter
\newcommand{\thickhline}{%
    \noalign {\ifnum 0=`}\fi \hrule height 1.5pt
    \futurelet \reserved@a \@xhline
}
\makeatother

\usepackage[margin=1in]{geometry}

\allowdisplaybreaks

\definecolor{mygreen}{RGB}{20,120,60}

\title{Stochastic Weighted Matching: $(1-\epsilon)$ Approximation}

\author{Soheil Behnezhad\thanks{Supported by a Google PhD Fellowship.} \and Mahsa Derakhshan}

\date{University of Maryland\\ \texttt{\{soheil,mahsa\}@cs.umd.edu}}

\newcommand{\local}[0]{\ensuremath{\mathsf{LOCAL}}}

\newcommand{\E}[0]{\ensuremath{\mathbb{E}}}

\DeclareMathOperator{\Var}{Var}
\DeclareMathOperator{\Cov}{Cov}

\newcommand{\opt}[0]{\ensuremath{\textsc{opt}}}

\newcommand{\gain}[1]{\ensuremath{g#1}}

\newcommand{\MM}[1]{\ensuremath{\mathsf{MM}(#1)}}
\newcommand{\cru}[1]{\ensuremath{G}}
\newcommand{\inm}[1]{\ensuremath{Z}}
\newcommand{\apxMM}[1]{\ensuremath{\mathsf{ApproxMatching}(#1)}}
\newcommand{\findmatching}[2]{\ensuremath{\mc{B}_{#1}( #2)}}
\SetKwRepeat{Do}{do}{while}

\renewcommand{\b}[1]{\ensuremath{\bm{\mathrm{#1}}}}
\DeclareMathOperator{\poly}{poly}
\DeclareMathOperator{\polylog}{polylog}

\renewcommand{\O}[1]{\ensuremath{O(#1)}}

\DeclareMathOperator{\var}{Var}

\renewcommand{\epsilon}[0]{\ensuremath{\varepsilon}}

\newcommand{\Bigmid}[0]{\ensuremath{\,\Big\vert\,}}

\let\originalleft\left
\let\originalright\right
\renewcommand{\left}{\mathopen{}\mathclose\bgroup\originalleft}
\renewcommand{\right}{\aftergroup\egroup\originalright}

\addauthor{sb}{blue}    
\addauthor{md}{red}    

\newtheorem{theorem}{Theorem}

\newtheorem{lemma}{Lemma}[section]

\newtheorem{proposition}[lemma]{Proposition}

\newtheorem{definition}[lemma]{Definition}
\newtheorem{claim}[lemma]{Claim}

\newtheorem{observation}[lemma]{Observation}
\newtheorem{remark}[lemma]{Remark}

\makeatletter
\def\thm@space@setup{%
  \thm@preskip= 0.2cm
  \thm@postskip=\thm@preskip 
}
\makeatother

\definecolor{mygreen}{RGB}{20,155,20}
\definecolor{myred}{RGB}{195,20,20}
\definecolor{linkcolor}{RGB}{0,0,230}
\definecolor{mylightgray}{RGB}{230,230,230}
\definecolor{verylightgray}{RGB}{240,240,240}
\definecolor{commentcolor}{RGB}{120,120,120}

\SetCommentSty{mycommfont}

\newcommand{\smparagraph}[1]{
\par\addvspace{0.2cm}
\noindent \textbf{#1}
}

\hypersetup{
     colorlinks=true,
     citecolor= mygreen,
     linkcolor= myred,
     urlcolor= mygreen
}

\newcommand{\etal}[0]{\textit{et al.}}

\newcommand{\mc}[1]{\ensuremath{\mathcal{#1}}}

\newcounter{myalgctr}

\newenvironment{tbox}{
\par\addvspace{0.2cm}
\begin{tcolorbox}[width=\textwidth,
                  enhanced,
                  boxsep=2pt,
                  left=1pt,
                  right=1pt,
                  top=4pt,
                  boxrule=1pt,
                  arc=0pt,
                  colback=white,
                  colframe=black,
                  unbreakable
                  ]
}{
\end{tcolorbox}
}

\newenvironment{tboxh}{
\par\addvspace{0.2cm}
\begin{tcolorbox}[width=\textwidth,
                  enhanced,
                  boxsep=2pt,
                  left=1pt,
                  right=1pt,
                  top=4pt,
                  boxrule=1pt,
                  arc=0pt,
                  colback=white,
                  colframe=black,
                  unbreakable,
                  float=t
                  ]
}{
\end{tcolorbox}
}

\newenvironment{graytbox}{
\par\addvspace{0.1cm}
\begin{tcolorbox}[width=\textwidth,
                  enhanced,
                  frame hidden,
                  boxsep=5pt,
                  left=1pt,
                  right=1pt,
                  top=2pt,
                  bottom=2pt,
                  boxrule=1pt,
                  arc=0pt,
                  colback=mylightgray,
                  colframe=black,
                  breakable
                  ]
}{
\end{tcolorbox}
}

\newcommand{\algcomment}[1]{{\color{gray} // #1 }}

\newcommand{\tboxhrule}[0]{\vspace{0.1cm} \hrule \vspace{0.2cm}}

\newenvironment{titledtbox}[1]{\begin{tbox}#1 \tboxhrule}{\end{tbox}}
\newenvironment{titledtboxh}[1]{\begin{tboxh}#1 \tboxhrule}{\end{tboxh}}

\newenvironment{tboxalg2e}[1]{
\refstepcounter{myalgctr}
	\begin{titledtbox}{\textbf{Algorithm \themyalgctr.} #1}
	\vspace{-0.2cm}
}
{
	\vspace{-0.3cm}
	\end{titledtbox}
}

\newcommand{\restatedesc}[3]{
\par\addvspace{0.2cm}
\noindent\textbf{#1 {\normalfont (#2)}.} {\em #3}
\par\addvspace{0.2cm}
}

\newcommand{\restate}[2]{
\restatedesc{#1}{restated}{#2}
}

\begin{document}
\maketitle

\thispagestyle{empty}
\begin{abstract}
\setlength{\parskip}{0.3em}
	Let $G=(V, E)$ be a given edge-weighted graph and let its {\em realization} $\mc{G}$ be a random subgraph of $G$ that includes each edge $e \in E$ independently with probability $p$. In the {\em stochastic matching} problem, the goal is to pick a sparse subgraph $Q$ of $G$ without knowing the realization $\mc{G}$, such that the maximum weight matching among the realized edges of $Q$ (i.e. graph $Q \cap \mc{G}$) in expectation approximates the maximum weight matching of the whole realization $\mc{G}$.
	
	In this paper, we prove that for any desirably small $\epsilon \in (0, 1)$, every graph $G$ has a subgraph $Q$ that guarantees a $(1-\epsilon)$-approximation and has maximum degree only $O_{\epsilon, p}(1)$. That is, the maximum degree of $Q$ depends only on $\epsilon$ and $p$ (both of which are known to be necessary) and not for example on the number of nodes in $G$, the edge-weights, etc.
	
	The stochastic matching problem has been studied extensively on both weighted and unweighted graphs. Previously, only existence of (close to) half-approximate subgraphs was known for weighted graphs [Yamaguchi and Maehara, SODA'18; Behnezhad~\etal{}, SODA'19]. Our result substantially improves over these works, matches the state-of-the-art for unweighted graphs [Behnezhad~\etal{}, STOC'20], and essentially settles the approximation factor.
\end{abstract}
\clearpage

{

\hypersetup{
     linkcolor= black
}

\thispagestyle{empty}
\tableofcontents{}
\clearpage
}

\setcounter{page}{1}

\section{Introduction}\label{sec:intro}

We study the {\em stochastic weighted matching} problem defined as follows. An arbitrary $n$-vertex graph $G=(V, E)$ with edge weights $w: E \to \mathbb{R}_{\geq 0}$ is given. A random subgraph $\mc{G}$ of $G$, called the {\em realization}, is then drawn by retaining each edge $e \in E$ independently with some fixed probability $p \in (0, 1]$. The goal is to choose a subgraph $Q$ of $G$ without knowing the realization $\mc{G}$ such that:
\begin{enumerate}[itemsep=0pt]
	\item The maximum weight matching (MWM) among the {\em realized} edges of $Q$ (i.e. graph $Q \cap \mc{G}$) approximates in expectation the MWM of the whole realization $\mc{G}$. Formally, we want the ``approximation factor'' $\E[\mu(Q \cap \mc{G})] / \E[\mu(\mc{G})]$ to be large where $\mu(\cdot)$ denotes the MWM's weight.
	\item The subgraph $Q$ has maximum degree $O(1)$. The constant here can (and in fact must) depend on $p$, but cannot depend on the structure of $G$ such as the number of nodes or edge-weights.
\end{enumerate}
Observe that by setting $Q = G$ we get an optimal solution, but the second constraint would be violated as the maximum degree in $G$ could be very large. On the other hand, if we choose $Q$ to be a single maximum weight matching of $G$, the maximum degree in $Q$ would desirably be only one, but it is not possible to guarantee anything better than a $p$-approximation for this algorithm\footnote{To see this, let $G$ be a clique with unit weights. It is easy to prove that a realization of $G$ has a near-perfect matching with high probability, whereas only $p$ fraction of the edges in the matching that forms $Q$ are realized.}. The stochastic matching problem therefore essentially asks whether it is possible to interpolate between these two extremes and pick a subgraph that is both sparse and provides a good approximation.

\smparagraph{Applications.} As its most straightforward application, the stochastic matching problem can be used as a {\em matching sparsifier} that approximately preserves the maximum (weight) matching under random edge failures \cite{sosa19}. It also has various applications in e.g. kidney exchange (see \cite{blumetalOR} for an extensive discussion) and online labor markets \cite{BR18,sosa19}. For these applications, one is only given the base graph $G$ but is tasked to find a matching in the realized subgraph $\mc{G}$. To do so, an algorithm can {\em query} each edge of $G$ to see whether it is realized. Each of these queries typically maps to a time-consuming operation such as interviewing a candidate and thus few queries must be conducted. To do so, one can (non-adaptively) query only the $O(n)$ edges of $Q$ and still expect to find an approximate MWM in the whole realization $\mc{G}$ which note may have $\Omega(n^2)$ edges.

\smparagraph{Known bounds.} As surveyed in Table~\ref{table:survey}, both the weighted and unweighted variants of this problem have been studied extensively \cite{blumetal,AKL16,AKL17,YM18,BR18,soda19,sosa19,sagt19,YM19, stoc20} since the pioneering work of Blum~\etal{}~\cite{blumetal}. For the unweighted case, earlier works achieved close to half approximation \cite{blumetal,AKL16, AKL17}. The second wave of results came close to $0.66$-approximation \cite{soda19,sosa19}. Eventually, it was shown in \cite{stoc20}  that the approximation factor can be made $(1-\epsilon)$ for any constant $\epsilon > 0$ \cite{stoc20}. All these results rely heavily on the underlying graph being unweighted. 

For the weighted case, in contrast, all known results remain close to half-approximation. The first result of this kind was proved by \cite{YM18} who showed that by allowing $Q$'s maximum degree to depend on the maximum weight $W$, one can obtain a $0.5$-approximation. It was later proved in \cite{BR18} through a different analysis of the same construction that dependence on $W$ is not necessary to achieve a $0.5$-approximation. Subsequently, the approximation factor was slightly improved to $0.501$ using a different construction \cite{soda19}. 


\newcommand{\STAB}[1]{\begin{tabular}{@{}c@{}}#1\end{tabular}}

\begin{table}[]
\centering
\begin{tabular}{|l|l|l|l|l|}
\hline
\rowcolor[HTML]{C0C0C0} 
{\color[HTML]{C0C0C0} }                          & {\color[HTML]{000000} Reference}                                                                         & {\color[HTML]{000000} Approx} & {\color[HTML]{000000} Degree of $Q$} & {\color[HTML]{000000} Notes} \\ \hline
                                                 & \parbox[t]{7.9cm}{Blum, Dickerson, Haghtalab, Procaccia, Sandholm, \& Sharma~\cite{blumetal,blumetalOR}} & $0.5 - \epsilon$                     & $O_{\epsilon, p}(1)$          &                              \\ \cline{2-5} 
                                                 & Assadi, Khanna, \& Li~\cite{AKL16}                                                                      & $0.5 - \epsilon$                     & $O_{\epsilon, p}(1)$          &                              \\ \cline{2-5} 
                                                 & Assadi, Khanna, \& Li~\cite{AKL17}                                                                      & $0.5001$                   & $O_{p}(1)$          &                              \\ \cline{2-5} 
                                                 & \parbox[t]{7.9cm}{Behnezhad, Farhadi, Hajiaghayi, \& Reyhani~\cite{soda19}}                              & $0.6568$                             & $O_{p}(1)$                    &                              \\ \cline{2-5} 
                                                 & Assadi \& Bernstein~\cite{sosa19}                                                                       & $2/3 - \epsilon$                     & $O_{\epsilon, p}(1)$          &                              \\ \cline{2-5} 
\multirow{-6}{*}{\STAB{\rotatebox[origin=c]{90}{Unweighted}}}                     & Behnezhad, Derakhshan, \& Hajiaghayi~\cite{stoc20}                                                      & $1-\epsilon$                         & $O_{\epsilon, p}(1)$          &                              \\ \thickhline
\multicolumn{1}{|c|}{}                           & Yamaguchi \& Maehara~\cite{YM18}                                                                        & $0.5 - \epsilon$                     & $O_{\epsilon, p}(W \log n)$ &                              \\ \cline{2-5} 
\multicolumn{1}{|c|}{}                           & Yamaguchi \& Maehara~\cite{YM18}                                                                        & $0.5 - \epsilon$                     & $O_{\epsilon, p}(W)$          & Bipartite                    \\ \cline{2-5} 
\multicolumn{1}{|c|}{}                           & Behnezhad \& Reyhani~\cite{BR18}                                                                       & $0.5 - \epsilon$                     & $O_{\epsilon, p}(1)$          &                              \\ \cline{2-5} 
\multicolumn{1}{|c|}{}                           & \parbox[t]{7.9cm}{Behnezhad, Farhadi, Hajiaghayi, \& Reyhani~\cite{soda19}}                              & $0.501$                   & $O_{p}(1)$                    &                              \\ \cline{2-5} 
\multirow{-5}{*}{\STAB{\rotatebox[origin=c]{90}{Weighted}}} & \textbf{This work}                                                                                       & $1-\epsilon$                         & $O_{\epsilon, p}(1)$          &                              \\ \hline
\end{tabular}
\caption{Survey of known results for weighted and unweighted graphs in chronological order. For simplicity we have hidden the actual dependence on $\epsilon$ and $p$ inside the $O$-notation. In the bounds above,  $W$ denotes the maximum edge-weight after scaling all the weights to integers.}
\label{table:survey}
\end{table}

\smparagraph{Our contribution.} Our main result in this paper is as follows:
\begin{graytbox}
	\begin{theorem}[Main result]\label{thm:main}
		For any weighted graph $G$, any $p \in (0, 1]$, and any $\epsilon > 0$, there is a construction of $Q$ with maximum degree $O_{\epsilon, p}(1)$ that guarantees a $(1-\epsilon)$-approximation for the weighted stochastic matching problem.
	\end{theorem}
\end{graytbox}

Not only Theorem~\ref{thm:main} is the first result showing that a significantly better than 0.5-approximation is achievable for weighted graphs, but it also essentially settles the approximation factor and the dependence of degrees on both $\epsilon$ and $p$ is necessary:

\begin{remark}\label{rem:apxoptimality}
	For any $\epsilon$, to obtain a $(1-\epsilon)$-approximation,  subgraph $Q$ should provably have maximum degree $\Omega(p^{-1} \log{\epsilon^{-1}})$ even when $G$ is a unit-weight clique \cite{AKL16}. This shows that dependence of degrees on both $\epsilon$ and $p$ is necessary, and the approximation factor cannot be made $(1-o(1))$ unless $Q$ has $\omega(1)$ degree.
\end{remark}

For simplicity of presentation, we do not calculate the precise dependence of the maximum degree of $Q$ on $\epsilon$ and $p$ in this paper. Though we remark that the $O_{\epsilon, p}(1)$ term in Theorem~\ref{thm:main} hides an exponential dependence on $\epsilon$ and $p$. We leave it as an open problem to determine whether a $\poly(\epsilon^{-1} p^{-1})$ degree subgraph can also achieve a $(1-\epsilon)$-approximation. 


\section{Technical Overview \& the Challenge with Weighted Graphs}\label{sec:techniques}

\newcommand{\sampling}[0]{\ensuremath{\mathsf{Sampling}}}

\newcommand{\greedy}[0]{\ensuremath{\mathsf{Greedy}}}

In the literature of the stochastic matching problem, the subgraph $Q$ typically has a very simple construction and much of the effort is concentrated on analyzing its approximation factor. A good starting point is the following \sampling{} algorithm proposed in \cite{soda19}:\footnote{As we will soon discuss, we do not analyze just the \sampling{} algorithm in this work, and combine it with a \greedy{} algorithm stated formally as  Algorithm~\ref{alg:complement}.} For some parameter $R = O_{\epsilon, p}(1)$, draw $R$ independent realizations $\mc{G}_1, \ldots, \mc{G}_R$ of $G$ and let $Q \gets \MM{\mc{G}_1} \cup \ldots \cup \MM{\mc{G}_R}$ where here $\MM{\cdot}$ returns a maximum weight matching. It is clear that the maximum degree of $Q$ is $R = O_{\epsilon, p}(1)$, but what approximation does it guarantee? Clearly $\E[\mu(\mc{G}_i)] = \E[\mu(\mc{G})]$ since each $\mc{G}_i$ is drawn from the same distribution as $\mc{G}$. However, observe that only $p$ fraction of the edges of each matching $\MM{\mc{G}_i}$ in expectation appear in the actual realization $\mc{G}$. Hence, the challenge in the analysis is to show that the realized edges of these matchings can augment each other to construct a matching whose weight approximates $\opt := \E[\mu(\mc{G})]$. 

Since the weighted stochastic matching problem is a generalization of the unweighted version, all the challenges that occur for the unweighted variant carry over to the weighted case. Of key importance, is the so called ``Ruzsa-Szemerédi barrier'' which was first observed by \cite{AKL16} toward achieving a $(1-\epsilon)$-approximation and was later broken in \cite{stoc20} for unweighted graphs using a notion of ``vertex independent matchings'' which we generalize to weighted graphs. Since the main contribution of this paper is solving the weighted version of the problem, we do not elaborate more on this barrier in this section and refer interested readers to Sections~1 and 2 of \cite{stoc20}. Instead, we discuss two challenges specific to weighted graphs and how we overcome them.



\smparagraph{Challenge 1: Low-probability/high-weight edges.} The analysis of the \sampling{} algorithm for unweighted graphs typically (see \cite{soda19,stoc20}) relies on a partitioning of the edge-set $E$ into ``crucial'' and ``non-crucial'' edges: Define $q_e := \Pr[e \in \MM{\mc{G}}]$ and let $\tau = \tau(\epsilon, p) \ll p$ be a sufficiently small threshold; an edge $e$ is called ``crucial'' if $q_e \geq \tau$ and ``non-crucial'' if $q_e < \tau$. Observe that if we draw say $R = \frac{\log 1/\epsilon}{\tau} = O_{\epsilon, p}(1)$ realizations in the \sampling{} algorithm, then nearly all crucial edges appear in at least one of $\MM{\mc{G}_1}, \ldots, \MM{\mc{G}_R}$ and thus belong to $Q$. On the other hand, non-crucial edges can be used very much interchangeably, at least when the graph is unweighted.

For weighted graphs there is a third class of edges: Edges $e$ with a small probability $q_e$ of appearing in $\MM{\mc{G}}$ but a relatively large weight $w_e$. On one hand, there could be a super-constant number of these edges connected to each vertex, so we cannot consider them crucial and add all of them to $Q$. On the other hand, even ``ignoring'' few edges of this type can significantly hurt the weight of the matching, so they cannot be regarded as non-crucial. This is precisely the reason that the analysis of \cite{soda19} only guarantees a $0.501$-approximation for weighted graphs but achieves up to $0.65$-approximation for unweighted graphs. (See  \cite[Section~6]{soda19} and in particular Figure~4 of \cite{soda19}.)

We handle low-probability/high-weight edges in a novel way. Particularly, we complement the \sampling{} algorithm (stated as Algorithm~\ref{alg:sampling}) with a \greedy{} algorithm (stated as Algorithm~\ref{alg:complement}) which hand picks {\em some} of the low-probability high-weight edges and adds them to $Q$. Then in our analysis, any low-probability/high-weight edge that is picked by the \greedy{} algorithm is treated as if they are crucial, while the rest are regarded as non-crucial. Describing how the \greedy{} algorithm decides which low-probability/high-weight edges to pick requires a number of careful definitions which are out of the scope of this section. However, in a rough sense, it picks edges that would be ``ignored''  in the analysis if we regarded them as non-crucial.

\smparagraph{Challenge 2: Lack of the ``sparsification lemma'' for weighted graphs.} Let us for now suppose that graph $G$ is unweighted. It is often useful to assume $\E[\mu(\mc{G})] = \Omega(n)$ as for instance  even by losing an additive $\epsilon n$ factor  in the size of the matching (say because a certain event fails around each vertex with probability $\epsilon$), we can still guarantee a multiplicative $(1-O(\epsilon))$-approximation.  A ``sparsification lemma'' of Assadi~\etal{}~\cite{AKL16} which was also used in a crucial way in \cite{stoc20} guarantees that this assumption comes without loss of generality for unweighted graphs. This is achieved by modifying the graph and ensuring that each vertex is matched with a large probability.

For weighted graphs, in contrast, the probability with which a vertex is matched is not a useful indicator of the weight that it contributes to the matching. For this reason, no equivalent of the sparsification lemma exists for weighted graphs. For another evidence that the sparsification lemma is not useful for weighted graphs, observe that by adding zero-weight edges we can assume w.l.o.g. that $G$ is a clique. Therefore, each vertex $v$ already has a probability $1-o(1)$ of being matched (but perhaps via a zero-weight edge) and thus the reduction of \cite{AKL16} does not help.

Due to lack of the sparsification lemma, it is not sufficient to simply bound the probability of a ``bad event'' around each vertex by say $\epsilon$ when the graph is weighted. Rather, it is important to analyze the actual expected loss to the weight conditioned on that this bad event occurs. For this reason, our analysis turns out to be much more involved than the unweighted case. This appears both in generalizing the vertex-independent lemma (Section~\ref{sec:vertexindependentlemma}) to the weighted case, and in various other places in the analysis (in particular Claims~\ref{cl:weightofg} and \ref{cl:xonNlarge}).

\section{Basic Definitions and The Algorithm}\label{sec:analysissetup}

\subsection{General Notation}\label{sec:prelim}

For any matching $M$, we use $w(M) := \sum_{e \in M} w_e$ to denote the weight of $M$; and use $v \in M$ for any vertex $v$ to indicate that there is an edge incident to $v$ that belongs to $M$. We use $\mu(H)$ to denote the weight of the maximum weight matching in graph $H$. For any two vertices $u$ and $v$, we use $d_G(u, v)$ to denote the size of the shortest path between $u$ and $v$ in graph $G$ (note that this is not their weighted distance). For any event $A$, we use $\pmb{1}(A)$ as the indicator of the event, i.e. $\pmb{1}(A) = 1$ if $A$ occurs and $\pmb{1}(A) = 0$ otherwise. 

\subsection{Basic Stochastic Matching Notation/Definitions}

We use $\opt$ to denote $\E[\mu(\mc{G})]$. Note that $\opt$ is just a number, the expected weight of the maximum weight matching in the realization $\mc{G}$. With this notation, to prove Theorem~\ref{thm:main}, we should prove that $\E[\mu(\mc{Q})] \geq (1-\epsilon) \opt$, where $\mc{Q} := Q \cap \mc{G}$ is the realized subgraph of $Q$.

For any graph $H$, we use $\MM{H}$ to denote a maximum weight matching of $H$. In case $H$ has multiple maximum weight matchings, $\MM{H}$ returns an arbitrary one. It would be useful to think of $\MM{\cdot}$ as a {\em deterministic} maximum weight matching algorithm that always returns the same matching for any specific input graph. Having this, for each edge $e$ define 
\begin{equation}\label{eq:defqandchi}
	q_e := \Pr_{\mc{G}}[e \in \MM{\mc{G}}] \qquad \text{and} \qquad \chi_e = w_e \cdot q_e.
\end{equation}
Observe that $\chi_e$ is the expected weight that $e$ contributes to matching $\MM{\mc{G}}$. These definitions also naturally extend to subsets of edges $F \subseteq E$ for which we denote
$$
	q(F) := \sum_{e \in F} q_e, \qquad \text{and} \qquad \chi(F) := \sum_{e \in F} \chi_e.
$$

\begin{observation}\label{obs:chiEisop}
	$\chi(E) = \opt$.
\end{observation}
\begin{proof}
	By definition $\opt = \E[\mu(\mc{G})]$. The proof therefore follows since:
	\begin{flalign*}
		\E[\mu(\mc{G})] &= \E[w(\MM{\mc{G}})] = \E\left[\sum_{e \in \MM{\mc{G}}} w_e\right] = \E\left[\sum_{e \in E} \pmb{1}(e \in \MM{\mc{G}}) \cdot w_e\right] = \sum_{e \in E} \Pr[e \in \MM{\mc{G}}] \cdot w_e\\
		&= \sum_{e \in E} q_e \cdot w_e = \sum_{e \in E} \chi_e = \chi(E),
	\end{flalign*}
	where the fourth equality follows simply from linearity of expectation.
\end{proof}

\subsection{The Algorithm}

\newcommand{\greedyalg}[1]{\ensuremath{\mathsf{GreedyAlgorithm}(#1)}}
\newcommand{\sampalg}[1]{\ensuremath{\mathsf{SamplingAlgorithm}(#1)}}

\newcommand{\initialncthresh}[0]{\ensuremath{\epsilon^5 p}}

In what follows we describe two different algorithms that each picks a subgraph of graph $G$. The final subgraph $Q$ is the union of the two subgraphs picked by these algorithms.

To state the first algorithm, let us first define function $\lambda: \mathbb{R} \times [0, 1] \to \mathbb{R}$ as:
\begin{equation}\label{eq:deflambda}
\lambda(\Delta, \epsilon) := \epsilon^{-24} (\log \Delta) (\log \log \Delta)^C,
\end{equation}
where $C \geq 1$ is a large enough absolute constants that we fix later. This perhaps strange-looking function is defined in this way so that it satisfies the various equations that we will need throughout the analysis. Having it, the first algorithm we use is as follows:

\begin{tboxalg2e}{\greedyalg{G=(V, E), p, \epsilon}}
\begin{algorithm}[H]
	\DontPrintSemicolon
	\SetAlgoSkip{bigskip}
	\SetAlgoInsideSkip{}
	
	\label{alg:complement}

	$P \gets \emptyset$.\;
	\While{true}{
		$\Delta \gets \max\{1, \text{maximum degree in subgraph $P$}\}$.\tcp*{So in the first iteration, $\Delta = 1$.}
		$I_q \gets \{ (u, v) \in E \setminus P \mid q_e \geq p^2 \epsilon^{10} \cdot \Delta^{-\lambda(\Delta, \epsilon)}  \}$.\;
		$I_d \gets \{ (u, v) \in E \setminus P \mid d_{P}(u, v) < \lambda(\Delta, \epsilon) \}$.\;
		$I \gets I_d \cup I_q$.\;\label{line:setI}
		\If{$\chi(I) \geq \epsilon \opt$\label{line:ifIlarge}}{
			$P \gets P \cup I$.\label{line:addtoQpp}
		}
		\Else {
			\Return $P$.
		}
	}
\end{algorithm}
\end{tboxalg2e}

From now on, when we use $\Delta$ we refer to the final value assigned to it during Algorithm~\ref{alg:complement}, which is equivalent to the maximum degree of $P$ (unless $P$ remains empty, which in that case $\Delta = 1$).

The second algorithm which was proposed first in \cite{soda19} is very simple and natural: Draw multiple random realizations and pick a maximum weight matching of each; formally:

\newcommand{\R}[0]{\ensuremath{p^{-2} \epsilon^{-10} \Delta^{\lambda(\Delta, \epsilon)}}}

\begin{tboxalg2e}{\sampalg{G=(V, E), p, \epsilon}}
\begin{algorithm}[H]
	\DontPrintSemicolon
	\SetAlgoSkip{bigskip}
	\SetAlgoInsideSkip{}
	
	\label{alg:sampling}

	$R \gets \lceil \R \rceil$.\;
	
	\For{$i$ in $1 \ldots R$}{
		Draw a realization $\mc{G}_i$ by retaining each edge $e \in E$ independently with probability $p$.\;
	}
	
	\Return $S := \MM{\mc{G}_1} \cup \ldots \cup \MM{\mc{G}_R}$.
\end{algorithm}
\end{tboxalg2e}

As mentioned earlier, the final subgraph $Q$ is the union of the outputs of Algorithms~\ref{alg:complement} and \ref{alg:sampling}. That is, $Q = S \cup P$.  We first prove in this section that the algorithms terminate and the resulting subgraph $Q$ has $O_{\epsilon, p}(1)$ maximum degree. We then turn to analyze the approximation-factor in the forthcoming sections.

\begin{lemma}
	Algorithms~\ref{alg:complement} and \ref{alg:sampling} terminate and the subgraph $Q$ has maximum degree $O_{\epsilon, p}(1)$.
\end{lemma}
\begin{proof}
	Algorithm~\ref{alg:complement} has an unconditional while loop, but we argue that it will terminate within at most $1/\epsilon$ iterations. To see this,  consider the progress of $\chi(P)$ after each iteration. Since none of the edges in $I$ are in $P$ due to its definition, in every iteration that the condition $\chi(I) \geq \epsilon \opt$ of Line~\ref{line:ifIlarge} holds, the value of $\chi(P)$ increases by at least $\epsilon \opt$. On the other hand, since $P \subseteq E$ and $\chi(E) = \opt$ (Observation~\ref{obs:chiEisop}), we have $\chi(P) \leq \opt$. Hence, after at most $1/\epsilon$ iterations, the condition of Line~\ref{line:ifIlarge} cannot continue to hold and the algorithm returns $P$. Algorithm~\ref{alg:sampling} also clearly terminates as it simply runs a for loop finitely many times.
	
	To bound the maximum degree of $Q$ by $O_{\epsilon, p}(1)$ we show that it suffices to bound the maximum degree $\Delta$ of $P$ by $O_{\epsilon, p}(1)$. To see this, first observe that if $\Delta = O_{\epsilon, p}(1)$ then also $\lambda(\Delta, \epsilon) = O_{\epsilon, p}(1)$ by definition of $\lambda$. On the other hand, since $S$ is simply the union of $R = O(\R)$ matchings, its maximum degree can also be bounded by $O(\R) = O_{\epsilon, p}(1)$. It thus only remains to prove $\Delta = O_{\epsilon, p}(1)$.
	
	To bound $\Delta$, let $\Delta_i$ be the maximum degree of $P$ by the end of iteration $i$ of the while loop in Algorithm~\ref{alg:complement}. We prove via induction that for any $i \leq 1/\epsilon$ we have $\Delta_i = O_{\epsilon, p}(1)$. This is sufficient for our purpose since we already showed above that the algorithm terminates within $1/\epsilon$ iterations.
	
	For the base case with $i = 0$ (i.e. before the start of the while loop) $P$ is empty, hence indeed $\Delta_0 = O_{\epsilon, p}(1)$. Now consider any iteration $i$. Take any vertex $v$ and let $e = (u, v)$ be an edge that belongs to $I$ at iteration $i$. By definition of $I$ in Line~\ref{line:setI}, $e \in I_d \cup I_q$ so it remains to bound the maximum degree of $I_d$ and $I_q$. If $e \in I_d$, there should be a path between $u$ and $v$ consisting of only the edges already in $P$ that has length less than $\ell := \lambda(\Delta_{i-1}, \epsilon)$. Since the maximum degree in $P$ at this point is $\Delta_{i-1}$, there are at most $\Delta_{i-1}^{\ell}$ such paths ending at $v$. This is a simple upper bound on the number of edges in $I_d$ connected to $v$ at iteration $i$. On the other hand, if $e \in I_q$, then by definition $q_e \geq p^2 \epsilon^{10} \cdot \Delta_{i-1}^{-\ell}$. Combined with $\sum_{e \ni v} q_e \leq 1$, this means there are at most $p^{-2} \epsilon^{-10} \cdot \Delta_{i-1}^{\ell}$ edges in $I_q$ connected to $v$. Thus the degree of any vertex $v$ increases by at most $ \Delta_{i-1}^{\ell} + p^{-2} \epsilon^{-10} \Delta_{i-1}^{\ell}$ and as a result:
	$$
		\Delta_i \leq \Delta_{i-1} + \Delta_{i-1}^{\ell} + p^{-2} \epsilon^{-10} \Delta_{i-1}^{\ell}.
	$$
	By the induction hypothesis, $\Delta_{i-1} = O_{\epsilon, p}(1)$ which also consequently implies $\ell = O_{\epsilon, p}(1)$ since $\ell$ is a function of only $\Delta_{i-1}$ and $\epsilon$. Therefore, $\Delta_i \leq O_{\epsilon, p}(1)^{O_{\epsilon, p}(1)} = O_{\epsilon, p}(1)$. Observe that since $i \leq 1/\epsilon$, this use of the asymptotic notation over the steps of the inductive argument does not lead to any undesirable blow-up and the final maximum degree is indeed $O_{\epsilon, p}(1)$ as desired.
\end{proof}

\section{The Analysis}\label{sec:analysis}

In this section, we analyze the approximation factor of the construction of $Q$ described in the previous section.

\smparagraph{Analysis via fractional matchings.} Recall that our goal is to show graph $\mc{Q} := Q \cap \mc{G}$ has a matching of weight $(1-O(\epsilon))\opt$ in expectation. Since $Q$ is constructed independently from the realization $\mc{G}$, one can think of $\mc{Q}$ as a subgraph of $Q$ that includes each edge of $Q$ independently with probability $p$. To show this subgraph $\mc{Q}$ has a matching of weight close to $\opt$, we follow the by now standard recipe \cite{soda19, stoc20}  of constructing a fractional matching $\b{x}$ on $\mc{Q}$, such that:
\begin{flalign}
	&& &x_v := \sum_{e \ni v} x_e \leq 1 && \forall v \in V &&  \label{eq:lpvertex} \\
	&& &x_e \geq 0 && \forall e \in \mc{Q} \label{eq:lpedge} \\
	&& &x(U) := \sum_{e = (u, v) : u, v \in U} x_e \leq \frac{|U| - 1}{2} && \forall U \subseteq V \text{ such that $|U|$ is odd and $\leq 1/\epsilon$. }\label{eq:lpblossom}
\end{flalign}
Here (\ref{eq:lpvertex}) and (\ref{eq:lpedge}) are simply fractional matching constraints. The last set of constraints (\ref{eq:lpblossom}), known as ``blossom'' \cite{edmonds1965maximum} constraints, are needed to ensure that our fractional matching $\b{x}$ can be turned into an integral matching of weight at least $(1-\epsilon)$ times that of $\b{x}$. (See \cite[Section~25.2]{schrijver2003combinatorial} for more context on the matching polytope and blossom constraints. See also \cite[Section~2.2]{soda19} for a simple proof of this folklore lemma that blossom inequalities over subsets of size up to $1/\epsilon$ are sufficient for a $(1-\epsilon)$-approximation.) In addition to the constraints above, we want fractional matching $\b{x}$ to have weight close to $\opt$ so that we can argue $\mc{Q}$ has an integral matching of size $(1-O(\epsilon))\opt$. Formally, our goal is to construct $\b{x}$ such that in addition to constraints (\ref{eq:lpvertex}--\ref{eq:lpblossom}), it satisfies:
\begin{equation}\label{eq:lpobjective}
	\E\left[\sum_{e \in \mc{Q}} x_e w_e\right] \geq (1-O(\epsilon))\opt.
\end{equation}

If $\b{x}$ satisfies all these constraints, then we have $\E[\mu(\mc{Q})] \geq (1-O(\epsilon)) \opt$, proving Theorem~\ref{thm:main}.

\begin{observation}\label{obs:towrapup}
	To prove Theorem~\ref{thm:main}, it suffices to give a construction $\b{x}: \mc{Q} \to [0, 1]$ satisfying constraints (\ref{eq:lpvertex}-\ref{eq:lpblossom}) and (\ref{eq:lpobjective}).
\end{observation}

The natural idea of using fractional matchings to analyze a solution for the stochastic matching problem was first used in \cite{soda19} and later in \cite{stoc20}. Among the two, only \cite{soda19} deals with weighted graphs, but there the  constructed fractional matching is only shown to have an expected weight of at least $0.501\opt$, guaranteeing only a $0.501$-approximation. Here, not only our subgraph $Q$ is constructed differently, but the way we construct the fractional matching is also fundamentally different and allows us to satisfy (\ref{eq:lpobjective}) and guarantee a $(1-\epsilon)$-approximation.

\subsection{Toward Constructing $x$: A Partitioning of $E$}

To construct fractional matching $\b{x}$, we first partition the edge set $E$ into $P \cup I' \cup N$, where $P$ is simply the output of Algorithm~\ref{alg:complement}, $I'$ is the set of edges in set $I$ defined in the last iteration of Algorithm~\ref{alg:complement} (for which the condition $\chi(I) \geq \epsilon \opt$ of Line~\ref{line:ifIlarge} fails), and $N$ is the rest of edges, i.e. $N = E - P - I'$. On all edges $e \in I'$ we simply set $x_e = 0$, i.e., we do not use them in the fractional matching $\b{x}$. For other edges $e \not\in I'$, we use different constructions for $\b{x}$ depending on whether $e \in P$ or $e \in N$. We describe the construction of $\b{x}$ on $P$ in Section~\ref{sec:xonP} and the construction on $N$ in Section~\ref{sec:xonN}. Before that, let us state a number of simple observations regarding this partitioning.

\begin{observation}\label{obs:chiP+chiNlarge}
	$\chi(P) + \chi(N) \geq (1-\epsilon)\opt$.
\end{observation}
\begin{proof}
	Recall that $\chi(E) = \opt$ by Observation~\ref{obs:chiEisop}. Combined with $E = P \cup I' \cup N$, this implies $\chi(P) + \chi(N) + \chi(I') \geq \opt$. To complete the proof, we argue that $\chi(I') \leq \epsilon \opt$. To see this, recall that $I'$ is defined as the set $I$ in the last iteration of Algorithm~\ref{alg:complement}. In the last iteration, the condition $\chi(I) \geq \epsilon \opt$ of Line~\ref{line:ifIlarge} in Algorithm~\ref{alg:complement} must fail (otherwise there would be another iteration), and thus $\chi(I') < \epsilon \opt$.
\end{proof}

\begin{observation}\label{obs:qesmallonNanddlarge}
	For any edge $e = (u, v) \in N$, $q_e < p^2 \epsilon^{10} \Delta^{-\lambda(\Delta, \epsilon)}$ and $d_P(u, v) \geq \lambda(\Delta, \epsilon)$.
\end{observation}
\begin{proof}
	In the last iteration of Algorithm~\ref{alg:complement}, all edges $e=(u, v)$ with $q_e \geq p^2 \epsilon^{10} \Delta^{-\lambda(\Delta, \epsilon)}$ or $d_P(u, v) < \lambda(\Delta, \epsilon)$ are either already in $P$ or are added to $I = I'$; thus $e \not\in N$ since $N = E - P - I'$. 
\end{proof}

\subsection{Construction of the Fractional Matching $x$ on $P$}\label{sec:xonP}

To describe the construction, let us first state a ``vertex-independent matching lemma'' which we will prove in Section~\ref{sec:vertexindependentlemma}.

\begin{lemma}\label{lemma:vertex-independent}
Let $G'=(V', E', w')$ be an edge-weighted base graph with maximum degree $\Delta'$. Let $\mc{G}'$ be a random subgraph of $G'$ that includes each edge $e \in E'$ independently with some probability $p \in (0, 1]$. Let $\mc{A}(H)$ be any (possibly randomized) algorithm that given any subgraph $H$ of $G'$, returns a (not necessarily maximum weight) matching of $H$. For any $\epsilon > 0$ there is a randomized algorithm $\mc{B}$ to construct a matching $\inm{} = \mc{B}(\mc{G}')$ of $\mc{G}'$ such that
\begin{enumerate}
\item For any vertex $v$, $\Pr_{\mc{G}'\sim G', \mc{B}}[v \in \inm{}] \leq \Pr_{\mc{G}'\sim G', \mc{A}}[v \in \mc{A}(\mc{G}')] + \epsilon^3.$
\item $\E[w(\inm{})] \geq (1-\epsilon)\E[w(\mc{A}(\mc{G}'))]$.
\item For any vertex-subset $\{v_1, v_2, \ldots \} \subseteq V'$ such that for all $i, j$, $d_{G'}(v_i, v_j) \geq \lambda$ where $\lambda = \O{\epsilon^{-24} \log \Delta' \poly(\log\log \Delta')}$, events $\{v_1 \in \inm{}\}, \{v_2 \in \inm{}\}, \{v_3 \in \inm{}\}, \ldots$ are all independent with respect to both the randomizations used in algorithm $\mc{B}$ and in drawing $\mc{G}'$.
\end{enumerate}
\end{lemma}


We use this lemma in the following way: The graph $G'=(V', E', w')$ of the lemma, is simply the subgraph $P$ picked by Algorithm~\ref{alg:complement} and thus $\Delta'$ is simply the maximum degree of $P$ which recall we denote by $\Delta$. We let the random subgraph $\mc{G}'$ be the subset of edges in $P$ that are realized, which we denote by $\mc{P}$. As discussed before, since $P$ is chosen independently from how the edges are realized, conditioned on $P$ each edge is still realized independently from the others, so the assumption that $\mc{P}$ is a random subgraph of $P$ with edges realized independently is valid. Finally, we define the algorithm $\mc{A}(H)$ of the lemma for any subgraph $H \subseteq P$ as follows:

\begin{tboxalg2e}{$\mc{A}(H)$}
\begin{algorithm}[H]
	\DontPrintSemicolon
	\SetAlgoSkip{bigskip}
	\SetAlgoInsideSkip{}
	
	\label{alg:A}

	$H' \gets H$.\;
	
	Add any edge $e \in E \setminus P$ independently with probability $p$ to $H'$.\;\label{line:AddtoHp}
	
	\Return $\MM{H'} \cap H$.
	
\end{algorithm}
\end{tboxalg2e}

Observe that with definition above, $\mc{A}(\mc{P})$ can be interpreted in the following useful way: The input subgraph $H = \mc{P}$ already includes each edge of $P$ independently with probability $p$. Since initially $H' \gets H$, and every edge $e \in E \setminus P$ is then added to $H'$ independently with probability $p$, by the end of Line~\ref{line:AddtoHp}, $H'$ will have the same distribution as the realization $\mc{G}$ of $G$. This means:

\begin{observation}\label{obs:distofAP}
	The output of $\mc{A}(\mc{P})$ has the same distribution as $\MM{\mc{G}} \cap P$.
\end{observation}

Finally, once we obtain a matching using the algorithm above, we remove each edge from the matching independently with probability $\epsilon$. Doing so, we only lose $\epsilon$ fraction of the weight of the matching in expectation, but we ensure that each vertex is matched with probability at most $1-\epsilon$ which will be useful later.

Let us for each vertex $v$ define $q_v^P := \sum_{e: v \in e, e \in P} q_e$ to be the probability that $v$ is matched in $\MM{\mc{G}}$ via an edge in $P$. Using Lemma~\ref{lemma:vertex-independent} as discussed above, we get:

\begin{claim}\label{cl:propsofZ}
	There is an algorithm $\mc{B}$ to construct a matching $Z$ on the realized edges $\mc{P}$ of $P$ s.t.:
	\begin{enumerate}
		\item For any vertex $v$, $\Pr_{\mc{P}, \mc{B}}[v \in \inm{}] \leq \min\{ q_v^P + \epsilon^3, 1-\epsilon \}.$
		\item $\E[w(\inm{})] \geq (1-2\epsilon)\chi(P)$.
		\item For any vertex-subset $\{v_1, v_2, \ldots \} \subseteq V$ such that for all $i, j$, $d_{P}(v_i, v_j) \geq \lambda(\Delta, \epsilon)$, events $\{v_1 \in \inm{}\}, \{v_2 \in \inm{}\}, \{v_3 \in \inm{}\}, \ldots$ are all independent with respect to both the randomizations used in algorithm $\mc{B}$ and the randomization in drawing $\mc{P}$.
		\item Matching $Z$ is independent of the realization of edges in $E \setminus P$.
	\end{enumerate}
\end{claim}
\begin{proof}
	For property~1, Lemma~\ref{lemma:vertex-independent} guarantees $\Pr_{\mc{P}, \mc{B}}[v \in Z] \leq \Pr_{\mc{P}, \mc{A}}[v \in \mc{A}(\mc{P})] + \epsilon^3$. Moreover,
	$$
	\Pr_{\mc{P}, \mc{A}}[v \in \mc{A}(\mc{P})] = \sum_{e \ni v} \Pr[e \in \mc{A}(\mc{P})] \stackrel{\text{Obs~\ref{obs:distofAP}}}{=} \sum_{e \ni v} \Pr[e \in \MM{\mc{G}} \cap P] = \sum_{e: v \in e, e \in P} \Pr[e \in \MM{\mc{G}}] =  q^P_v.
	$$
	Therefore, $\Pr[v \in Z] \leq q^P_v + \epsilon^3$. On the other hand, since as discussed above, at the end we drop each edge from the matching independently with probability $\epsilon$, $\Pr[v \in Z] \leq 1-\epsilon$. Combination of these two bounds proves property~1.
	
	For property~2, Lemma~\ref{lemma:vertex-independent} already guarantees that the reported matching has weight at least $(1-\epsilon)\E[w(\mc{A}(\mc{P}))]$. Since on top of that we retain each edge of the final matching with probability $1-\epsilon$, we lose another $(1-\epsilon)$ factor and have $\E[w(Z)] \geq (1-2\epsilon)\E[w(\mc{A}(\mc{P}))]$. To see why this is the claimed bound of property~2, observe that:
	$$
		\E[w(\mc{A}(\mc{P}))] \stackrel{\text{Obs~\ref{obs:distofAP}}}{=} \E[w(\MM{\mc{G}} \cap P)] = \sum_{e \in P} \Pr[e \in \MM{\mc{G}}] w_e = \sum_{e \in P} \chi_e = \chi(P).
	$$
	
	For property~3, it just suffices to make sure $\lambda(\Delta, \epsilon) \geq \lambda$ where recall $\lambda(\Delta, \epsilon)$ was defined in (\ref{eq:deflambda}) whereas $\lambda$ is defined in Lemma~\ref{lemma:vertex-independent}. By definition (\ref{eq:deflambda}), we already have $\lambda(\Delta, \epsilon) = \Omega(\lambda)$. On the other hand, in definition (\ref{eq:deflambda}) of $\lambda(\Delta, \epsilon)$ there is a constant $C$ that we can tune. Picking this constant to be large enough, we can guarantee that $\lambda(\Delta, \epsilon) \geq \lambda$ and satisfy this property.
		
	Finally, property~4 holds since in construction of $Z$ the algorithm is essentially unaware of the actual realization of edges in $E \setminus P$ and is thus independent of it.
\end{proof}

Once we construct matching $Z$ on the realized edges of $P$ using the algorithm above, for any edge $e \in P$ we set $x_e = 1$ if $e \in Z$ and $x_e = 0$ otherwise. Therefore, $\b{x}$ is in fact integral on all edges of $P$. The properties of $Z$ highlighted in Claim~\ref{cl:propsofZ} will be later used in augmenting $\b{x}$ via the realized edges among the edges in $N$.

\subsection{Construction of the Fractional Matching $x$ on $N$}\label{sec:xonN}

We first formally describe construction of $\b{x}$ on the edges in $N$, then discuss the main intuitions behind the construction, and finally prove that it satisfies the needed properties.

\subsubsection{The Construction}

We first define an ``assignment'' $\b{f}: E \to [0, 1]$, then based on $\b{f}$ define an assignment $\b{g}: E \to [0, 1]$, then based on $\b{g}$ define an assignment $\b{h}: E \to [0, 1]$, and finally construct $\b{x}$ from $\b{h}$. For any assignment $\b{a} \in \{\b{f}, \b{g}, \b{h}, \b{x}\}$ we may use the following notation: For an edge $e$, $a_e$ denotes the value of  $\b{a}$ on edge $e$. For a vertex $v$, $a_v := \sum_{e \ni v} a_e$ denotes the sum of assignments adjacent to $v$. The weight $w(\b{a})$ denotes $\sum_{e \in E} a_e w_e$.

As outlined above, we first define $\b{f}: E \to [0, 1]$ on each edge $e$ as follows:
\begin{equation}\label{eq:def-f}
	f_e := \begin{cases}
		\frac{1}{R} \sum_{i=1}^R \pmb{1}(e \in \MM{\mc{G}_i}) & \text{if $e \in N$},\\
		0 & \text{otherwise,}
	\end{cases}
\end{equation}
where recall that $\mc{G}_i$ is the $i$th drawn realization in Algorithm~\ref{alg:sampling} and $R$ is the total number of realizations drawn in Algorithm~\ref{alg:sampling}. In words, for any $e \in N$, the value of $f_e$ denotes the fraction of matchings $\MM{\mc{G}_1}, \ldots, \MM{\mc{G}_R}$ that include $e$.

Based on $\b{f}$, we define $\b{g}$ on each $e = (u, v)$ as:
\begin{equation}\label{eq:def-g}
	g_e := \begin{cases}
		f_e & \text{if $f_e \leq p^2 \epsilon^7 \Delta^{-\lambda(\Delta, \epsilon)}$, $f_u \leq 1-q^P_u + \epsilon^3$, and $f_v \leq 1- q^P_v + \epsilon^3$,}\\
		0 & \text{otherwise.}
	\end{cases}
\end{equation}

Next, based on $\b{g}$, we define $\b{h}$ on each edge $e = (u, v)$ as:
\begin{equation}\label{eq:def-h}
	h_e := \begin{cases}	
	\frac{g_e}{p \Pr[v \not\in Z] \Pr[u \not\in Z]} & \text{if $u \not\in Z$, $v \not\in Z$, and $e$ is realized}\\
	0 & \text{otherwise.}
	\end{cases}
\end{equation}
Here, as defined in the previous section, the value of $q_v^P$ for a vertex $v$ denotes the probability that $v$ is matched in $\MM{\mc{G}}$ via an edge in $P$.

We are finally ready to define the construction of $\b{x}$ on $N$. On each edge $e=(u, v) \in N$, we set:
\begin{equation}\label{eq:def-x-onEP}
	x_e \gets \begin{cases}
		\frac{h_e}{1+3\epsilon} & \text{if $h_v \leq 1+3\epsilon$ and $h_u \leq 1+3\epsilon$,}\\
		0 & \text{otherwise.}
	\end{cases}
\end{equation}

\subsubsection{Intuitions and Proof Outline}

Here we discuss the main intuitions behind the construction above for $\b{x}$ on $N$ in a slightly informal way. The rigorous proof that the final fractional matching $\b{x}$ satisfies properties (\ref{eq:lpvertex}-\ref{eq:lpobjective}) is given in the forthcoming sections.

As mentioned above, for every edge $e \in N$, $f_e$ simply denotes the fraction of matchings $\MM{\mc{G}_1}, \ldots, \MM{\mc{G}_R}$ that include $e$. Therefore $\b{f}$ is a linear combination of these integral matchings, and thus is a valid fractional matching. Another key observation here is that since each $\mc{G}_i$ has the same distribution as $\mc{G}$, the probability of each edge $e$ appearing in each matching $\MM{\mc{G}_i}$ is exactly equal to the probability $q_e$ that it appears in $\MM{\mc{G}}$. This can be used to prove $\E[f_e] = q_e$ (see Observation~\ref{obs:Efeqe}) which also implies $\E[w(\b{f})] = \chi(N)$ (see Observation~\ref{obs:weightoff}). Thus, fractional matching $\b{f}$ has precisely the weight $\chi(N)$ we need $\b{x}$ to have on $N$. In addition (unlike $q_e$) the value of $f_e$ is only non-zero on edges $e \in N$ that also belong to the output $S$ of Algorithm~\ref{alg:sampling}. This is desirable since recall that if an edge $e \in N$ does not belong to $S$, then $e \not\in Q$ and as a result $e \not\in \mc{Q}$. Thus, we should ensure $x_e = 0$ since we want $\b{x}$ to be a fractional matching of subgraph $\mc{Q}$.

In the next step of the construction, we define $\b{g}$ based on $\b{f}$. The key idea behind this definition is to get rid of possible ``deviations'' in $\b{f}$ and ensure that $\b{g}$ satisfies certain deterministic inequalities for $g_e$ on all edges $e$, and $g_v$ for all vertices $v$. It turns out that by carefully bounding the probability of these deviations, we can still argue that $\b{g}$ has weight close to $\chi(N)$ (see Claim~\ref{cl:weightofg}) just like $\b{f}$.

Despite the desirable properties mentioned above, $\b{g}$ is still far from the values we would like to assign to edges $N$ in $\b{x}$, for the following two reasons. First, we want $\b{x}$ to be non-zero only on $\mc{Q}$, i.e. the realized edges in $Q$. However, in defining $\b{g}$ we never look at edge realizations. Hence, it could be that $g_e > 0$ for an edge $e$ that is not realized. The second problem is that we need to augment the matching $Z$ already constructed in Section~\ref{sec:xonP}. More specifically, recall from Section~\ref{sec:xonP} that we have already assigned $x_e = 1$ to any edge $e \in Z$. Therefore, if we want $\b{x}$ to be a valid fractional matching, all edges $e$ that are incident to a matched vertex of $Z$ should have $x_e = 0$. In defining $\b{h}$, we address both issues at the same time. That is, for any edge $e$, if $e$ is not realized or at least one of its endpoints is matched in $Z$, we set $h_e = 0$. Though note that we still want $\E[w(\b{h})]$ to be close to $\E[w(\b{g})]$ and $\chi(N)$. To compensate for the loss to the weight due to edges $e$ for which $g_e > 0$ but $h_e = 0$, on each edge $e$ that is eligible to be assigned $h_e > 0$, we multiply $g_e$ by an appropriate amount that cancels out the probability of assigning $h_e = 0$. Doing so, we can ensure that $\E[w(\b{h})]$ remains sufficiently close to $w(\b{g})$ and thus $\chi(N)$ (Claim~\ref{cl:hlarge}). 

Finally, recall from above that $\b{f}$ is a valid fractional matching and thus so is $\b{g}$ since $g_e \leq f_e$ on all edges. A next challenge is to make sure that once we obtain $\b{h}$ by multiplying $\b{g}$ on some edges, we still have a valid fractional matching. That, e.g. $h_v \leq 1$ for all vertices $v$. Toward achieving this, we first show in Claim~\ref{cl:prhvlargegivenFe} that for each vertex $v$, the probability that $h_v > 1+3\epsilon$ is very small. But these deviations do occur. Thus, in our final construction of $\b{x}$, on any edge $e = (u, v)$ for which at least one of $h_u$ and $h_v$ exceeds $1+3\epsilon$, we set $x_e = 0$ and set $x_e = h_e/(1+3\epsilon)$ on the rest of the edges. This way, we guarantee that for any vertex $v$, $x_v \leq 1$. Moreover, due to the low probability of violations in $\b{h}$, there is a small probability for any edge $e$ to have $x_e = 0$ but $h_e > 0$. Therefore, $\b{x}$ as defined, will have weight close to $\chi(N)$ in expectation on the edges in $\mc{Q} \cap N$ (Claim~\ref{cl:xonNlarge}). Combined with the construction of $\b{x}$ on the edges in $P$ which guarantees a weight of $\approx \chi(P)$ there, we obtain that overall $\b{x}$ will have weight close to $\chi(P) + \chi(N)$ which is $\approx \opt$ as guaranteed by Observation~\ref{obs:chiP+chiNlarge}. Therefore, $\b{x}$ can be shown to satisfy all the needed properties required by Observation~\ref{obs:towrapup} thereby proving Theorem~\ref{thm:main} (see Section~\ref{sec:wrapup}).

\subsubsection{Properties of $f$ and $g$.}

We start with a few simple observations.

\begin{observation}\label{obs:preMMiisq}
	For any $i \in [R]$ and any edge $e$, $\Pr[e \in \MM{\mc{G}_i}] = q_e$.
\end{observation}
\begin{proof}
	Since each realization $\mc{G}_i$ in Algorithm~\ref{alg:sampling} has the same distribution as $\mc{G}$, we have $\Pr[e \in \MM{\mc{G}_i}] = \Pr[e \in \MM{\mc{G}}]$. The claim follows from the definition (\ref{eq:defqandchi}) that $\Pr[e \in \MM{\mc{G}}] = q_e$.
\end{proof}

\begin{observation}\label{obs:Efeqe}
	For each edge $e \in N$, $\E[f_e] = q_e$.
\end{observation}
\begin{proof}
	For any $e \in N$, it holds by definition (\ref{eq:def-f}) that
$$
	\E[f_e] = \frac{1}{R} \sum_{i=1}^R \Pr[e \in \MM{\mc{G}_i}] \stackrel{\text{Obs}~\ref{obs:preMMiisq}}{=} \frac{1}{R} \sum_{i=1}^R q_e = q_e,
$$
which is the desired bound.
\end{proof}

\begin{observation}\label{obs:weightoff}
	$\E[w(\b{f})] = \chi(N)$.
\end{observation}
\begin{proof}
 We have $w(\b{f}) = \sum_{e \in E} f_e w_e = \sum_{e \in N}f_e w_e$ since $f_e = 0$ for all $e \not\in N$. Thus by linearity of expectation,
$$
\E[w(\b{f})] = \sum_{e \in N} \E[f_e] w_e = \sum_{e \in N} q_e w_e = \chi(N),
$$
where the second equality holds by Observation~\ref{obs:Efeqe}.
\end{proof}

\begin{observation}\label{obs:gesmall}
	For any edge $e$, $g_e \leq p^2 \epsilon^7 \Delta^{-\lambda(\Delta, \epsilon)}$.
\end{observation}
\begin{proof}
	By construction of $\b{g}$, if $g_e$ is non-zero, then $g_e = f_e$ and $f_e \leq p^2 \epsilon^7 \Delta^{-\lambda(\Delta, \epsilon)}$.
\end{proof}

\begin{observation}\label{obs:gvsmall}
	For any vertex $v$, $g_v \leq 1-q^P_u + \epsilon^3$.
\end{observation}
\begin{proof}
	By construction of $\b{g}$, if $g_v \not= 0$, then $f_v \leq 1 - q^P_u + \epsilon^3$, and thus so is $g_v$ since $\b{g} \leq \b{f}$.
\end{proof}

The main takeaway of this section is the following claim, which guarantees $\E[w(\b{g})]$ is large enough for our purpose. 

\begin{claim}\label{cl:weightofg}
	$\E[w(\b{g})] \geq (1-\epsilon)\chi(N)$.
\end{claim}

The proof of Claim~\ref{cl:weightofg} is rather involved. The main difficulty is the lack of an equivalent of a sparsification lemma for weighted graphs (as discussed in Section~\ref{sec:techniques}). The rest of this section is devoted to proving Claim~\ref{cl:weightofg} for which we need a number of other auxiliary claims.

For simplicity, let us for each edge $e$ use $F_e$ as a shorthand for event $f_e \leq p^2 \epsilon^{7} \Delta^{-\lambda(\Delta, \epsilon)}$ and for each vertex $v$ use $F_v$ as a shorthand for event $f_v \leq 1-q^P_v + \epsilon^3$. These are precisely the events used in definition (\ref{eq:def-g}) of $\b{g}$. In particular, for any $e = (u, v) \in E$, $g_e = f_e$ if event $F_e \wedge F_u \wedge F_v$ holds.

\begin{claim}\label{cl:Egunionbound}
	For any edge $e \in N$,
	$$
	\E[g_e] \geq q_e (1-\Pr[\overline{F_e} \mid \mc{G}_1] - \Pr[\overline{F_u} \mid \mc{G}_1] - \Pr[\overline{F_v} \mid \mc{G}_1]),
	$$
	where here as usual, $\overline{F_e}, \overline{F_v}, $ and $\overline{F_u}$ denote the complement of events $F_e$, $F_v$, and $F_u$ respectively.
\end{claim}

\begin{proof}	
	We have
	\begin{flalign*}
		&& \E[g_e] &= \E[f_e \mid F_e \wedge F_u \wedge F_v]\\
		&& &= \E\left[ \frac{1}{R} \sum_{i=1}^R \pmb{1}(e \in \MM{\mc{G}_i}) \,\Big\vert\, F_e \wedge F_u \wedge F_v \right] && \text{By definition (\ref{eq:def-f}) and noting $e \in N$.}\\
		&& &= \frac{1}{R} \sum_{i=1}^R \Pr[e \in \MM{\mc{G}_i} \mid F_e \wedge F_u \wedge F_v ] && \text{Linearity of expectation.}\\
		&& &= \frac{1}{R} \sum_{i=1}^R \Pr[e \in \MM{\mc{G}_1} \mid F_e \wedge F_u \wedge F_v ] && \text{By symmetry.}\\
		&& &= \Pr[e \in \MM{\mc{G}_1} \mid F_e \wedge F_u \wedge F_v ]\\
		&& &= \Pr[e \in \MM{\mc{G}_1}] \cdot \frac{\Pr[F_e \wedge F_u \wedge F_v \mid \mc{G}_1]}{\Pr[F_e \wedge F_v \wedge F_u]} && \text{Bayes' rule.}\\
		&& &\geq \Pr[e \in \MM{\mc{G}_1}] \cdot \Pr[F_e \wedge F_u \wedge F_v \mid \mc{G}_1] && \text{Since $\Pr[F_e \wedge F_v \wedge F_u] \leq 1$.}\\
		&& &= q_e \Pr[F_e \wedge F_u \wedge F_v \mid \mc{G}_1] &&\text{By Observation~\ref{obs:preMMiisq}.}\\
		&& &\geq q_e (1-\Pr[\overline{F_e} \mid \mc{G}_1] - \Pr[\overline{F_u} \mid \mc{G}_1] - \Pr[\overline{F_v} \mid \mc{G}_1]). && \text{By union bound.}
	\end{flalign*}
	The last inequality matches the one stated in the claim and the proof is complete.
\end{proof}

\begin{claim}\label{cl:boundfe}
	For any edge $e \in N$, it holds that $\Pr[\overline{F_e} \mid \mc{G}_1] \leq 2\epsilon^3$.
\end{claim}

\begin{proof}
	We have
	\begin{flalign*}
		\E[f_e \mid \mc{G}_1] &= \E\left[\frac{1}{R} \sum_{i=1}^R \pmb{1}(e \in \MM{\mc{G}_i}) \,\Big|\, \mc{G}_1 \right] \leq \frac{1}{R} + \frac{1}{R} \sum_{i=2}^R \Pr[e \in \MM{\mc{G}_i}] \stackrel{\text{Obs}~\ref{obs:preMMiisq}}{\leq} \frac{1}{R} + q_e.
	\end{flalign*}
	We have $R \geq p^{-2} \epsilon^{-10} \Delta^{\lambda(\Delta, \epsilon)}$ by its definition in Algorithm~\ref{alg:sampling} and also $q_e \leq p^2 \epsilon^{10} \Delta^{-\lambda(\Delta, \epsilon)}$ by Observation~\ref{obs:qesmallonNanddlarge}. Hence, $\E[f_e \mid \mc{G}_1] < 2 p^2 \epsilon^{10} \Delta^{-\lambda(\Delta, \epsilon)}$. Applying Markov's inequality, we thus get
	$$
		\Pr\left[f_e > p^2 \epsilon^7 \Delta^{-\lambda(\Delta, \epsilon)} \mid \mc{G}_1 \right] = \Pr[\overline{F_e} \mid \mc{G}_1] \leq 2\epsilon^3,
	$$
	which is the desired bound.
\end{proof}

\begin{claim}\label{cl:boundfv}
	For any vertex $v$, $\Pr[\overline{F_v} \mid \mc{G}_1] \leq 4\epsilon^4$.
\end{claim}
\begin{proof}
	Let us for any $i \in [R]$ define $X_i = 1$ if vertex $v$ is matched in $\MM{\mc{G}_i}$ via an edge $e \in N$ and $X_i = 0$ otherwise. Also let $X := \sum_{i=2}^R X_i$ (note that the sum index starts from 2). We have:
	\begin{flalign}
		\nonumber f_v &= \sum_{e \ni v} f_e  = \sum_{e: v \in e, e \in N} f_e \qquad\qquad\qquad \text{By (\ref{eq:def-f}), $f_e = 0$ if $e \not\in N$.}\\
		\nonumber &= \sum_{e: v \in e, e \in N} \Bigg(\frac{1}{R} \sum_{i=1}^R \pmb{1}(e \in \MM{\mc{G}_i}) \Bigg) = \frac{1}{R} \sum_{i=1}^R \sum_{e: v \in e, e \in N} \pmb{1}(e \in \MM{\mc{G}_i}) \\
		&= \frac{1}{R} \sum_{i=1}^R X_i \leq  \frac{1}{R} + \frac{1}{R} \sum_{i=2}^R X_i \leq \frac{X + 1}{R}. \label{eq:hmDEUgGCL}
	\end{flalign}
	Furthermore,
	\begin{flalign}
		\nonumber && \Pr[\overline{F_v} \mid \mc{G}_1] &= \Pr[f_v > 1 - q^P_v + \epsilon^3 \mid \mc{G}_1] && \text{Definition of $F_v$.}\\
		\nonumber && &\leq \Pr\left[\frac{X+1}{R} > 1 - q^P_v + \epsilon^3 \mid \mc{G}_1 \right] && \text{By (\ref{eq:hmDEUgGCL}), $f_v \leq \frac{X + 1}{R}$.}\\
		\nonumber && &= \Pr\left[X > R(1 - q^P_v + \epsilon^3) - 1 \mid \mc{G}_1 \right]\\
		&& &= \Pr\left[X > R(1 - q^P_v + \epsilon^3) - 1\right], \label{eq:ueTHEsu}
	\end{flalign}
	where the last inequality follows from the fact that $X = \sum_{i=2}^R X_i$ depends only on realizations $\mc{G}_2, \ldots, \mc{G}_R$ and is independent of realization $\mc{G}_1$.
	
	Therefore to bound $\Pr[\overline{F_v} \mid \mc{G}_1]$ we should analyze the behavior of random variable $X$. Let us start with its expected value:
	\begin{flalign}
	\nonumber \E\left[X\right] &= \sum_{i=2}^R \Pr[X_i = 1] = \sum_{i=2}^R \Pr[X_2 = 1] && \text{As by symmetry $\Pr[X_2 = 1] = \ldots = \Pr[X_R = 1]$.} \\
	\nonumber &= (R-1) \Pr[X_2 = 1] \leq R \Pr[X_2 = 1] \\
	& \leq R(1- q^P_v).\label{eq:muGlhuc3}
	\end{flalign}
	The last inequality holds for the following reason: By definition $q^P_v = \sum_{e: v \in e, e \in P} q_e$; since each edge $e$ belongs to $\MM{\mc{G}_2}$ with probability $q_e$ by Observation~\ref{obs:preMMiisq}, we get that with probability $q^P_v$, vertex $v$ is matched in $\MM{\mc{G}_2}$ via an edge $e \in P$; in this case, event $X_2 = 1$ which requires $v$ to be matched via an edge in $N$ cannot hold since $N \cap P = \emptyset$; hence $\Pr[X_2 = 1] \leq 1 - q^P_v$.
	
	We also need a concentration bound on $X$ which we prove via Chebyshev's inequality\footnote{One can also attempt to get a stronger concentration bound via Chernoff-type bounds, but the second moment method suffices for our purpose here.} using the independence of events $X_2, \ldots, X_R$. For any $t \geq 0$ we have
	\begin{equation}\label{eq:GcreLtsmx}
		\Pr[X > \E[X] + t] \leq \frac{\Var[X]}{t^2} = \frac{\sum_{i=2}^R \Var[X_i]}{t^2} \leq \frac{R \Var[X_2]}{t^2} = \frac{R(\E[X_2^2] - E[X_2]^2)}{t^2} \leq \frac{R}{t^2}.
	\end{equation}
	As a result,
	\begin{equation}\label{eq:jhteEhc}
		\Pr[X > R(1-q^P_v + \epsilon^3) - 1] = \Pr[X > R(1-q^P_v) + (\epsilon^3 R - 1)] \stackrel{(\ref{eq:muGlhuc3}), \, (\ref{eq:GcreLtsmx})}{\leq} \frac{R}{(\epsilon^3 R - 1)^2} \leq 4\epsilon^4,
	\end{equation}
	where the last inequality follows from 
	$$
		\frac{R}{(\epsilon^3 R - 1)^2} \leq \frac{R}{(\epsilon^3 R / 2)^2} \leq \frac{4}{\epsilon^6 R} \stackrel{R \geq p^{-2} \epsilon^{-10}}{\leq} \frac{4 \epsilon^{10}p^2}{\epsilon^6}  \leq 4\epsilon^4.
	$$
	Replacing (\ref{eq:jhteEhc}) into (\ref{eq:ueTHEsu}) gives the desired bound that $\Pr[\overline{F_v} \mid \mc{G}_1] \leq 4 \epsilon^4$.
\end{proof}

We finally have the tools needed to prove Claim~\ref{cl:weightofg}.

\begin{proof}[Proof of Claim~\ref{cl:weightofg}]
	We have
	\begin{equation}\label{eq:hucolcehuastlll}
		\E[w(\b{g})] = \E\left[\sum_{e \in E} g_e w_e \right] \geq \E\left[\sum_{e \in N} g_e w_e \right] = \sum_{e \in N} \E[g_e] w_e.
	\end{equation}
	Furthermore, by Claim~\ref{cl:Egunionbound}, for any $e \in N$ we have
	$$
		\E[g_e] \geq q_e (1-\Pr[\overline{F_e} \mid \mc{G}_1] - \Pr[\overline{F_u} \mid \mc{G}_1] - \Pr[\overline{F_v} \mid \mc{G}_1]).
	$$
	Incorporating the bounds of Claims~\ref{cl:boundfe} and \ref{cl:boundfv}, we get for any $e \in N$ that
	$$
		\E[g_e] \geq q_e (1 - 2 \epsilon^3 - 4 \epsilon^4 - 4 \epsilon^4) > (1-10\epsilon^3) q_e.
	$$
	Therefore, from (\ref{eq:hucolcehuastlll}) we get
	$$
		\E[w(\b{g})] \geq \sum_{e \in N} (1-10\epsilon^3)q_e w_e = (1-10\epsilon^3) \sum_{e \in N} q_e w_e = (1-10\epsilon^3) \chi(N) \geq (1-\epsilon) \chi(N),
	$$
	concluding the proof.
\end{proof}

\subsubsection{Properties of $h$, and $x$ on $N$.}

In this section we turn to prove a number of useful properties of $\b{h}$. We emphasize that in the previous section all expectations and probabilities are taken only over the randomization inherent in Algorithm~\ref{alg:sampling}. In contrast, in this section, all the probabilistic statements are with regards to the randomization of realization $\mc{G}$, and the randomization used in drawing matching $Z$ in Section~\ref{sec:xonP}.

\begin{claim}\label{cl:hlarge}
	$\E[w(\b{h})] \geq w(\b{g})$.
\end{claim}
\begin{proof}
	Take any edge $e = (u, v) \in N$. By definition of \b{h} we have $h_e = \frac{g_e}{p \Pr[v \not\in Z] \Pr[u \not\in Z]}$ if $e$ is realized and both $u$ and $v$ are unmatched in $Z$, and $h_e = 0$ otherwise. Since $d_P(u, v) \geq \lambda(\Delta, \epsilon)$ by Observation~\ref{obs:qesmallonNanddlarge}, the condition of Claim~\ref{cl:propsofZ} part 3 is satisfied and events $u \in Z$ and $v \in Z$ are independent. Moreover, since $e \not\in P$, its realization is also independent of $Z$ by Claim~\ref{cl:propsofZ} property~4. Hence,
	$$
		\E[h_e] = \Pr[\text{$e$ realized}] \Pr[v \not\in Z] \Pr[u \not\in Z] \frac{g_e}{p \Pr[v \not\in Z] \Pr[u \not\in Z]} = g_e.
	$$
	This means that 
	$$
	\E[w(\b{h})] = \sum_{e \in N} \E[h_e] w_e = \sum_{e \in N} g_e w_e = w(\b{g}),
	$$
	completing the proof.
\end{proof}

\begin{observation}\label{obs:hesmall}
	For any edge $e$, $h_e \leq \frac{g_e}{p \epsilon^2} \leq p \epsilon^5 \Delta^{-\lambda(\Delta, \epsilon)}$.
\end{observation}
\begin{proof}
	By construction of $\b{h}$ for any $e = (u, v)$ we have
	$$
		h_e \leq \frac{g_e}{p \Pr[v \not\in Z] \Pr[u \not\in Z]} \stackrel{\star}{\leq} \frac{g_e}{p\epsilon^2} \stackrel{\text{Observation~\ref{obs:gesmall}}}{\leq}  \frac{p^2 \epsilon^7 \Delta^{-\lambda(\Delta, \epsilon)}}{p \epsilon^2} = p \epsilon^5 \Delta^{-\lambda(\Delta, \epsilon)},
	$$
	where the inequality marked by $\star$ follows from the fact that $\Pr[v \in Z] \leq 1- \epsilon$ by property~1 of Claim~\ref{cl:propsofZ} and thus $\Pr[v \not\in Z] \geq \epsilon$ and similarly $\Pr[u \not\in Z] \geq \epsilon$.
\end{proof}

Claim~\ref{cl:xonNlarge} below is one of the key components towards achieving our main result in Theorem~\ref{thm:main}. We present the proof in multiple steps, by proving a number of properties of $\b{h}$.

\begin{claim}\label{cl:xonNlarge}
	It holds that $\E[\sum_{e \in N} x_e w_e] \geq (1-15\epsilon) w(\b{g})$.
\end{claim}
\begin{proof}
	We already know from Claim~\ref{cl:hlarge} that $\E[w(\b{h})] \geq w(\b{g})$. Thus, if we show  $\E[\sum_{e \in N} x_e w_e] \geq (1-3\epsilon) \E[w(\b{h})]$ we are done. For brevity, for any edge $e = (u, v)$ we use $H_{e}$ to indicate the event $(u \not\in Z, v \not\in Z, e \text{ realized})$. Also we use $X_e$ to indicate event $(h_v \leq 1+3\epsilon \text{ and } h_u \leq 1+\epsilon)$. Observe that $H_e$ is the event used in construction (\ref{eq:def-h}) of $h_e$ and $X_e$ is the event used in construction (\ref{eq:def-x-onEP}) of $\b{x}$ on $N$. Putting together (\ref{eq:def-h}) and (\ref{eq:def-x-onEP}), for any $e = (u,v) \in N$, we have
	$$
		x_e = \begin{cases}
			\frac{1}{1+3\epsilon} \cdot \frac{g_e}{p \Pr[u \not\in Z] \Pr[v \not\in Z]} & H_e \wedge X_e,\\
			0 & \text{otherwise.}
		\end{cases}
	$$
	This means that
	\begin{flalign*}
		\E\Bigg[\sum_{e \in N} x_e w_e\Bigg] &= \sum_{e \in N} \E[x_e] w_e\\
		&= \sum_{e \in N} \Pr[H_e \wedge X_e] \frac{1}{1+3\epsilon} \cdot \frac{g_e}{p \Pr[u \not\in Z] \Pr[v \not\in Z]} w_e\\
		&= \frac{1}{1+3\epsilon} \sum_{e \in N} \Pr[X_e \mid H_e] \Pr[H_e] \frac{g_e}{p \Pr[u \not\in Z] \Pr[v \not\in Z]} w_e\\
		&= \frac{1}{1+3\epsilon} \sum_{e \in N} \Pr[X_e \mid H_e] \E[h_e] w_e\\
		&= \frac{1}{1+3\epsilon} \sum_{e=(u, v) \in N} \Pr[h_v \leq 1 + 3\epsilon \wedge h_u \leq 1 + 3\epsilon \mid H_e] \E[h_e] w_e\\
		&= \frac{1}{1+3\epsilon} \sum_{e=(u, v) \in N} (1 - \Pr[h_v > 1 + 3\epsilon \mid H_e] - \Pr[h_u > 1 + 3\epsilon \mid H_e]) \E[h_e] w_e.
	\end{flalign*}
	Therefore it only remains to bound $\Pr[h_v > 1+3 \epsilon \mid H_e]$. The following claim, whose proof we present after the proof of the current Claim~\ref{cl:xonNlarge}, gives us the desired bound for it.
	
	\begin{claim}\label{cl:prhvlargegivenFe}
		Let edge $e = (u, v) \in N$ be the one fixed above, then  
		$\Pr_{\mc{G},Z}[h_{v} > 1+3\epsilon \mid F_e] \leq 6\epsilon.
		$
	\end{claim}
	
	Plugging Claim~\ref{cl:prhvlargegivenFe} this into the equation above, we thus get
	$$
		\E\Bigg[\sum_{e \in N} x_e w_e\Bigg] \geq \frac{1-12 \epsilon}{1+3\epsilon} \sum_{e \in N} \E[h_e] w_e = \frac{1-12 \epsilon}{1+3\epsilon} \E[w(\b{h})] > (1-15 \epsilon) \E[w(\b{h})] \stackrel{\text{Claim~\ref{cl:hlarge}}}{\geq} (1-15\epsilon) w(\b{g}),
	$$
	which is our desired bound.
\end{proof}

For the rest of this section, we fix $e = (u, v) \in N$ and focus on proving  Claim~\ref{cl:prhvlargegivenFe}. To do so, we first bound the expected value of $h_v$ conditioned on $H_e$ in Claim~\ref{cl:exphv} and then finish the proof via a concentration bound.
 
 Note from constructions (\ref{eq:def-f}), (\ref{eq:def-g}), and (\ref{eq:def-h}) of respectively $\b{f}$, $\b{g}$, and $\b{h}$, that $h_{e'} = g_{e'} = f_{e'} = 0$ for any $e' \not\in N$. Hence, we have
	$
		h_v = \sum_{e' \ni v} h_{e'} = \sum_{e': e' \in N, v \in e'} h_{e'}.
	$
	Now let $e_1=(v, u_1), e_2 = (v, u_2), \ldots, e_k=(v, u_k)$ be all edges connected to vertex $v$ that belong to $N$ and assume that $e_1 = e = (v, u)$. We thus have
	\begin{equation}\label{eq:hvissumhei}
	h_v = \sum_{i=1}^k h_{e_i}.
	\end{equation}
	
	\begin{claim}\label{cl:exphv}
		Let edge $e = (u, v) \in N$ be the one fixed above, then $\E[h_v \mid H_e] \leq 1+2\epsilon$.
	\end{claim}
	\begin{proof}
	We have
	\begin{equation}\label{eq:hUceLcejm}
		\E[h_v \mid H_e] = \E\left[ \sum_{i=1}^k h_{e_i} \Bigmid H_e \right] = \sum_{i=1}^k \E[h_{e_i} \mid H_e].
	\end{equation}
	To bound this, consider the following partitioning of $\{e_1, \ldots, e_k\}$ into two subsets $A$ and $B$:
	$$
		A = \{ e_i \mid d_P(u_i, u) < \lambda(\Delta, \epsilon) \}, \qquad B = \{ e_i \mid d_P(u_i, u) \geq \lambda(\Delta, \epsilon) \}.
	$$
	In particular, observe that $e_1 \in A$ since $u_1 = u$ which implies $d_P(u_1, u) = 0$. Separating $A$ and $B$ in the sum of (\ref{eq:hUceLcejm}) we get
	\begin{equation}\label{eq:Xxublhaesateus}
		\E[h_v \mid H_e] = \sum_{e_i \in A} \E[h_{e_i} \mid H_e] + \sum_{e_i \in B} \E[h_{e_i} \mid H_e].
	\end{equation}
	We bound the two sums over $A$ and $B$ in the inequality above separately.
	
	\smparagraph{Bounding the sum over $A$.} For each $h_{e_i} \in A$, we use the pessimistic upper bound of Observation~\ref{obs:hesmall} for $h_{e_i}$. But instead we bound the size of $A$ by 
	\begin{equation}\label{eq:sizeofA}
		|A| \leq \Delta^{\lambda(\Delta, \epsilon)} + 1 \leq 2 \Delta^{\lambda(\Delta, \epsilon)}.
	\end{equation}
	This first inequality follows from the fact that the maximum degree in $P$ is bounded by $\Delta$, and hence there are at most $\Delta^{\lambda(\Delta, \epsilon)}$ nodes (other than $u$ itself) that have distance less than $\lambda(\Delta, \epsilon)$ to $u$ in graph $P$. The second inequality simply follows from the fact that both $\Delta$ and $\lambda(\Delta, \epsilon)$ are $\geq 1$ (see Algorithm~\ref{alg:complement}). We thus have 
	\begin{flalign}
		&& \nonumber \sum_{e_i \in A} h_{e_i} &\leq p \epsilon^5 \Delta^{-\lambda(\Delta, \epsilon)} |A| && \text{By Observation~\ref{obs:hesmall}}\\
		&& &\leq 2 p\epsilon^5. && \text{By (\ref{eq:sizeofA}).}\label{eq:sumforA}
	\end{flalign}
	
	\smparagraph{Bounding the sum over $B$.} Recall that $H_e = (e \text{ realized}, v \not\in Z, u \not\in Z)$ and $H_{e_i} = (e_i \text{ realized}, v \not\in Z, u_i \not\in Z)$. Therefore for any edge $e_i \in B$, we have
	\begin{flalign*}
		\Pr[H_{e_i} \mid H_e] &= \Pr[e_i \text{ realized}, v \not\in Z, u_i \not\in Z \mid e \text{ realized}, v \not\in Z, u \not\in Z]\\
		&= \Pr[e_i \text{ realized}, u_i \not\in Z \mid e \text{ realized}, v \not\in Z, u \not\in Z]\\
		&= p \Pr[u_i \not\in Z \mid e \text{ realized}, v \not\in Z, u \not\in Z]\\
		&= p \Pr[u_i \not\in Z \mid v \not\in Z, u \not\in Z],
	\end{flalign*}
	where the last two equalities follow from property~4 of Claim~\ref{cl:propsofZ} regarding independence of matching $Z$ from realization of edges in $N$ (such as $e_i$ and $e$), and noting that $e_i \not= e$ since $e_i \in B$. On the other hand, since $d_P(u_i, u) \geq \lambda(\Delta, \epsilon)$ based on definition of $B$, and $d_P(u_i, v) \geq \lambda(\Delta, \epsilon)$ by Observation~\ref{obs:qesmallonNanddlarge}, we get that event $u_i \in Z$ is independent of $v \in Z, u \in Z$ due to property~3 of Claim~\ref{cl:propsofZ}. Therefore $\Pr[u_i \not\in Z \mid v \not\in Z, u \not\in Z] = \Pr[u_i \not\in Z]$ and thus
	\begin{equation}\label{eq:aCethhaaau}
		\Pr[H_{e_i} \mid H_e] = p \Pr[u_i \not\in Z] \qquad\qquad \text{for any } e_i \in B.
	\end{equation}
	We can therefore bound the sum in (\ref{eq:Xxublhaesateus}) over $B$ as follows:
	\begin{flalign*}
		\sum_{e_i \in B} \E[h_{e_i} \mid H_e] &= 		\sum_{e_i \in B} \frac{g_{e_i} \Pr[H_{e_i} \mid H_e]}{p\Pr[v \not\in Z]\Pr[u_i \not\in Z]}\\
		&= \sum_{e_i \in B} \frac{g_{e_i}}{\Pr[v \not\in Z]} && \text{By (\ref{eq:aCethhaaau}).}\\
		&\leq \frac{g_v}{\Pr[v \not\in Z]}\\
		&\leq \frac{1-q^P_v + \epsilon^3}{\Pr[v \not\in Z]} && \text{Observation~\ref{obs:gvsmall}.}\\
		&\leq \frac{1-q^P_v + \epsilon^3}{1 - \min\{q^P_v + \epsilon^3, 1-\epsilon\}}. && \parbox{6.5cm}{Since $\Pr[v \in Z] \leq \min\{q^P_v + \epsilon^3, 1-\epsilon\}$ by Claim~\ref{cl:propsofZ}.}
	\end{flalign*}
	Since both the nominator and the denominator are $\approx 1 - q^P_v$, the sum is upper bounded by $\approx 1$. To formalize this, consider two scenarios: (i) $q^P_v - \epsilon^3 \geq 1-\epsilon$, and (ii) $q^P_v - \epsilon^3 < 1- \epsilon$. In the former, we have
	$$
		\frac{1-q^P_v + \epsilon^3}{1 - \min\{q^P_v + \epsilon^3, 1-\epsilon\}} \stackrel{(i)}{=} \frac{1-q^P_v + \epsilon^3}{1 - (1-\epsilon)} \stackrel{(i)}{\leq} \frac{1 - (1-\epsilon + \epsilon^3) + \epsilon^3}{\epsilon} = \frac{\epsilon}{\epsilon} = 1.
	$$
	In the latter case,
	$$
		\frac{1-q^P_v + \epsilon^3}{1 - \min\{q^P_v + \epsilon^3, 1-\epsilon\}} \stackrel{(ii)}{=} \frac{1-q^P_v + \epsilon^3}{1 - q^P_v - \epsilon^3} \leq \frac{1-(1-\epsilon + \epsilon^3)+\epsilon^3}{1-(1-\epsilon + \epsilon^3)-\epsilon^3} = \frac{\epsilon}{\epsilon(1 - 2\epsilon^2)} \leq 1 + \epsilon,
	$$
	where the last inequality holds for any $\epsilon < 0.36$. Therefore overall, we get
	\begin{equation}\label{eq:sumforB}
		\sum_{e_i \in B} \E[h_{e_i} \mid H_e] \leq 1 + \epsilon.
	\end{equation}
	
	Incorporating the bounds (\ref{eq:sumforA}) and (\ref{eq:sumforB}) into (\ref{eq:Xxublhaesateus}) we get that
	$\E[h_v \mid H_e] \leq 1 + \epsilon + p \epsilon^5 \leq 1+2\epsilon$.
\end{proof}

We are now ready to prove Claim~\ref{cl:prhvlargegivenFe} via a concentration bound.

\begin{proof}[Proof of Claim~\ref{cl:prhvlargegivenFe}]
	By Chebyshev's inequality, and the bound $\E[h_v \mid H_e] \leq 1 + 2\epsilon$ of Claim~\ref{cl:exphv}, we get that
	\begin{equation}\label{eq:hvchebyshev}
		\Pr_{\mc{G},Z}[h_v > (1 + 2\epsilon) + \epsilon \mid H_e] \leq \frac{\Var_{\mc{G},Z}[h_v \mid H_e]}{\epsilon^2}.
	\end{equation}
	For brevity, we do not write the subscript $\mc{G}, Z$ for our probabilistic statements for the rest of the proof when it is clear. Since $h_v = \sum_{i=1}^k h_{e_i}$, by definition of variance we have
	$$
		\Var[h_v \mid H_e] = \sum_{i=1}^k \sum_{j=1}^k \Cov[h_{e_i}, h_{e_j} \mid H_e].
	$$
	By definition, if $h_{e_i}$ and $h_{e_j}$ are independent with respect to the randomization of $\mc{G}$ and $Z$, and conditioned on $H_e$, then $\Cov_{\mc{G},Z}[h_{e_i}, h_{e_j} \mid H_e] = 0$. But this does not hold for all $h_{e_i}$ and $h_{e_j}$. As in the proof of Claim~\ref{cl:exphv} consider the following partitioning of $\{e_1, \ldots, e_k\}$:
	$$
		A = \{ e_i \mid d_P(u_i, u) < \lambda(\Delta, \epsilon) \}, \qquad B = \{ e_i \mid d_P(u_i, u) \geq \lambda(\Delta, \epsilon) \}.
	$$
	With this partitioning, we can rewrite the equation above for variance as:
	\begin{flalign}
	\nonumber \Var[h_v \mid H_e] &= \sum_{e_i \in A} \sum_{j=1}^k  \Cov[h_{e_i}h_{e_j} \mid H_e] + \sum_{e_i \in B} \sum_{e_j \in A}  \Cov[h_{e_i}h_{e_j} \mid H_e] + \sum_{e_i \in B} \sum_{e_j \in B} \Cov[h_{e_i}h_{e_j} \mid H_e]\\
	&\leq 2 \sum_{e_i \in A} \sum_{j=1}^k  |\Cov[h_{e_i}h_{e_j} \mid H_e]| + \sum_{e_i \in B} \sum_{e_j \in B} \Cov[h_{e_i}h_{e_j} \mid H_e].\label{eq:varisumsofAB}
	\end{flalign}
	We will bound the two sums over $A$ differently. Before that, let us prove a simple upper bound on the covariance of any two edges $e_i, e_j$:
	\begin{flalign}
		\nonumber && \Cov[h_{e_i} h_{e_j} \mid H_e] &= \E_{\mc{G},Z}[h_{e_i} h_{e_j} \mid H_e] - \E_{\mc{G},Z}[h_{e_i} \mid H_e] \E[h_{e_j} \mid H_e]\\
		\nonumber && &\leq \E_{\mc{G},Z}[h_{e_i} h_{e_j} \mid H_e]\\
		&& &\leq \frac{g_{e_i}}{p \epsilon^2}\cdot \frac{g_{e_j}}{p \epsilon^2} && \text{By Observation~\ref{obs:hesmall}.}\label{eq:covariancebound}
	\end{flalign}
	
	\smparagraph{Bounding the sums over $A$.} We have
	\begin{flalign}
		\nonumber 2 \sum_{e_i \in A} \sum_{j=1}^k  |\Cov[h_{e_i}h_{e_j} \mid H_e]| &\leq 2 \sum_{e_i \in A} \sum_{j=1}^k  \frac{g_{e_i}}{p \epsilon^2}\cdot \frac{g_{e_j}}{p \epsilon^2} && \text{By (\ref{eq:covariancebound}).}\\
		\nonumber &\leq 2 \sum_{e_i \in A} \sum_{j=1}^k  \epsilon^3 \Delta^{-\lambda(\Delta, \epsilon)} g_{e_j} && \text{$g_{e_i} \leq p^2 \epsilon^7 \Delta^{-\lambda(\Delta, \epsilon)}$ by Observation~\ref{obs:gesmall}.}\\
		\nonumber &= 2 \epsilon^3 \Delta^{-\lambda(\Delta, \epsilon)} \sum_{e_i \in A} \sum_{j=1}^k g_{e_j}\\
		\nonumber &= 2 \epsilon^3 \Delta^{-\lambda(\Delta, \epsilon)} |A| g_v\\
		\nonumber &\leq 2 \epsilon^3 \Delta^{-\lambda(\Delta, \epsilon)} |A| && \text{Since $\b{g}$ is a valid fractional matching.}\\
		&\leq 4 \epsilon^3. && \text{Since $|A| \leq 2\Delta^{\lambda(\Delta, \epsilon)}$ by (\ref{eq:sizeofA}).}\label{eq:covforA}
	\end{flalign}
	
	\smparagraph{Bounding the sum over $B$.}  Let us for each $e_i \in B$ use $D_i$ to denote the set of edges $e_j \in B$ where $\Cov[h_{e_i}, h_{e_j} \mid H_e] \not= 0$. We claim that for each $e_i \in B$, $|D_i| \leq \Delta^{\lambda(\Delta, \epsilon)}$. To prove this, observe that for all $e_i, e_j \in B$, we have $d_P(u_i, u) \geq \lambda(\Delta, \epsilon)$ and $d_P(u_j, u) \geq \lambda(\Delta, \epsilon)$ by definition of $B$. Moreover, since $(u, v), (v, u_i), (v, u_j) \in N$, we have $d_P(u, v) \geq \lambda(\Delta, \epsilon)$, $d_P(u_i, v) \geq \lambda(\Delta, \epsilon)$, and $d_P(u_j, v) \geq \lambda(\Delta, \epsilon)$ by Observation~\ref{obs:qesmallonNanddlarge}. Therefore among $\{v, u, u_i, u_j\}$ only the pair $u_i, u_j$ may have $d_P(u_i, u_j) < \lambda(\Delta, \epsilon)$. If this is not the case and $d_P(u_i, u_j) \geq \lambda(\Delta, \epsilon)$, then based on Claim~\ref{cl:propsofZ}  events $H_{e_i}$ and $H_{e_j}$, and consequently, $h_{e_i}$ and $h_{e_j}$ would be independent conditioned on $H_e$ and thus $\Cov(h_{e_i}, h_{e_j} \mid H_e) = 0$. This means that indeed for any $e_i$ and any $e_j \in D_i$, $d_P(u_i, u_j) \leq \lambda(\Delta, \epsilon)$. Since the maximum degree of $P$ is $\Delta$, there are at most $\Delta^{\lambda(\Delta, \epsilon)}$ such vertices, implying indeed that
	\begin{equation}\label{eq:jjCletaou}
		|D_i| \leq \Delta^{\lambda(\Delta, \epsilon)} + 1 \stackrel{\text{(\ref{eq:sizeofA})}}{\leq} 2 \Delta^{\lambda(\Delta, \epsilon)} \qquad \text{for any $e_i \in B$.}
	\end{equation}
	We therefore have:
	\begin{flalign}
		\nonumber \sum_{e_i \in B} \sum_{e_j \in B} \Cov[h_{e_i}h_{e_j} \mid H_e] &= \sum_{e_i \in B} \sum_{e_j \in D_i } \Cov[h_{e_i}h_{e_j} \mid H_e]\\
		\nonumber &\leq \sum_{e_i \in B} \sum_{e_j \in D_i } \frac{g_{e_i}}{p\epsilon^2} \frac{g_{e_j}}{p \epsilon^2} && \text{By (\ref{eq:covariancebound}).}\\
		\nonumber &\leq \frac{1}{p^2 \epsilon^4} \sum_{e_i \in B} g_{e_i} \Bigg(\sum_{e_j \in D_i } g_{e_j}\Bigg)\\
		\nonumber &\leq \frac{1}{p^2 \epsilon^4} \sum_{e_i \in B} g_{e_i} \Bigg(\sum_{e_j \in D_i } p^2 \epsilon^7 \Delta^{-\lambda(\Delta, \epsilon)} \Bigg) && \text{By Observation~\ref{obs:gesmall}.}\\
		\nonumber &\leq \frac{p^2 \epsilon^7 \Delta^{-\lambda(\Delta, \epsilon)}}{p^2 \epsilon^4} \sum_{e_i \in B} g_{e_i} |D_i|\\
		\nonumber &\leq 2\epsilon^3 \sum_{e_i \in B} g_{e_i} && \text{By (\ref{eq:jjCletaou}) $|D_i| \leq 2\Delta^{\lambda(\Delta, \epsilon)}$.}\\
		&\leq 2 \epsilon^3 g_v \leq 2 \epsilon^3. &&\parbox{4.5cm}{Since $\b{g}$ is a valid fractional matching.}\label{eq:covforB}
	\end{flalign}
	Incorporating (\ref{eq:covforA}) and (\ref{eq:covforB}) into (\ref{eq:varisumsofAB}) we get that $\Var[h_v \mid H_e] \leq 4\epsilon^3 + 2\epsilon^3 = 6 \epsilon^3$. Replacing back to equation (\ref{eq:hvchebyshev}) we get that $\Pr[h_v > 1+3\epsilon \mid H_e] \leq 6\epsilon^3/\epsilon^2 = 6\epsilon$.
\end{proof}

\subsection{Putting Everything Together}\label{sec:wrapup}

In this section we prove using the stated bounds above that $\b{x}$ as constructed satisfies the fractional matching constraints (\ref{eq:lpvertex}-\ref{eq:lpblossom}), satisfies (\ref{eq:lpobjective}), i.e. has expected weight at least $(1-O(\epsilon))\opt$, and that it is non-zero only on the edges of $\mc{Q}$. This as already described in Observation~\ref{obs:towrapup} completes the proof of Theorem~\ref{thm:main} that subgraph $Q$ guarantees a $(1-\epsilon)$-approximation.

\smparagraph{Fractional matching constraints (\ref{eq:lpvertex}) and (\ref{eq:lpedge}).} For constraint (\ref{eq:lpvertex}) that $x_v \leq 1$ for any vertex $v$, consider two scenarios: If $v$ is matched via a matching edge of $Z$ (the matching constructed in Section~\ref{sec:xonP} on $P$), then on all edges $e \in N$ we set $x_e = 0$ by construction of $\b{h}$ (\ref{eq:def-h}) and thus $x_v = 1$. On the other hand, if $v$ is unmatched in $Z$, then we still have $x_v \leq 1$ due to construction (\ref{eq:def-x-onEP}) of $\b{x}$ based on $\b{h}$ which guarantees $\b{x} \leq \frac{1}{1+3\epsilon} \b{h}$ and in addition $x_v = 0$ if $h_v \geq 1 + 3\epsilon$.

The constraint (\ref{eq:lpedge}) that $x_e \geq 0$ for all edges $e$ is easy to confirm. For edges in $P$, the value of $x_e$ is either 0 or 1. For edges in $N$, since $\b{f}$ is non-negative, so are $\b{g}$, $\b{h}$, and $\b{x}$. 

\smparagraph{Blossom inequalities (\ref{eq:lpblossom}).} The blossom constraint (\ref{eq:lpblossom}) that $x(U) \leq \frac{|U|-1}{2}$ for all odd size $U \subseteq V$ with $|U| \leq 1/\epsilon$ follows for the following reason. There are two types of edges that form $\b{x}$ by construction: Those in set $P$, and those in $N$. For any edge $e \in P$, the value of $x_e$ is simply integral. For any $e \in N$, we have
\begin{equation}\label{eq:buahoeudlcDDD}
	x_e \stackrel{(\ref{eq:def-x-onEP})}{<} h_e \stackrel{\text{Observation~\ref{obs:hesmall}}}{\leq} p \epsilon^5 \Delta^{-\lambda(\Delta, \epsilon)} \leq p \epsilon^5 \leq \epsilon^5.
\end{equation}
Now suppose for contradiction that there is a subset of size $\leq 1/\epsilon$ for which the blossom constraint (\ref{eq:lpblossom}) is violated, and let $U$ be the smallest such subset. If there is an edge $e = (u, v) \in P$ whose both endpoints are in $U$ and $x_e = 1$, then one can confirm that subset $U \setminus \{u, v\}$ should also violate the blossom inequality contradicting that $U$ is the smallest. On the other hand, for all edges $e$ with both endpoints in $U$ we have $x_e \leq \epsilon^5$ by (\ref{eq:buahoeudlcDDD}). Since there are at most $|U|^2$ edges inside $U$ and $|U| \leq 1/\epsilon$, we have $x(U) \leq |U|^2 \epsilon^5 \leq \epsilon^{-2} \epsilon^5 = \epsilon^{3} < 1 < \frac{|U|-1}{2}$, contradicting the fact that the blossom inequality is violated. So all blossom inequalities of size up to $1/\epsilon$ must be satisfied.

\smparagraph{Fractional matching $\b{x}$ is non-zero only on $\mc{Q}$.} For any edge $e \in P$, if $x_e > 0$ then $e \in Z$ and by Claim~\ref{cl:propsofZ}, $e \in \mc{P}$ i.e. $e$ is realized. Since $P \subseteq Q$, then $e \in \mc{Q}$. On the other hand, for any edge $e \in N$, if $x_e > 0$ then we should have $h_e > 0$ by construction of $\b{x}$ and to have $h_e > 0$ we should have $g_e > 0$ and $f_e > 0$. By construction of $\b{h}$, if $h_e > 0$ then $e$ must be realized, and by construction of $\b{f}$, if $f_e > 0$ then $e \in S \subseteq Q$. Combination of these imply $e \in \mc{Q}$. Therefore overall, if for any edge $e$, $x_e > 0$ then $e \in \mc{Q}$ and so $\b{x}$ is a fractional matching of only the edges in $\mc{Q}$. 

\smparagraph{Expected weight of $\b{x}$.} By Claim~\ref{cl:propsofZ} part 2, we have $\E[w(Z)] \geq (1-2\epsilon) \chi(P)$ and thus $\E[\sum_{e \in P} x_e w_e] \geq (1-2\epsilon) \chi(P)$. On the other hand, by Claim~\ref{cl:xonNlarge} $\E[\sum_{e \in N} x_e w_e] \geq (1-15\epsilon) w(\b{g})$ and $\E[w(\b{g})] \geq (1-\epsilon)\chi(N)$ by Claim~\ref{cl:weightofg}. Combining all of these, we get
\begin{flalign*}
	\E[w(\b{x})] &= \E\left[ \sum_{e \in E} x_e w_e \right] = \E\left[ \sum_{e \in P} x_e w_e \right] + \E\left[ \sum_{e \in N} x_e w_e \right] \geq (1-2\epsilon) \chi(P) + (1-15\epsilon)(1-\epsilon) \chi(N)\\
	&\geq (1-16\epsilon) (\chi(P) + \chi(N)) \stackrel{\text{Obs~\ref{obs:chiP+chiNlarge}}}{\geq} (1-16 \epsilon)(1-\epsilon)\opt \geq (1-17\epsilon)\opt.
\end{flalign*}
And thus our construction of $\b{x}$ satisfies $\E[w(\b{x})] \geq (1-O(\epsilon)) \opt$ required by (\ref{eq:lpobjective}).

Combination of the properties above as shown before in Observation~\ref{obs:towrapup} proves Theorem~\ref{thm:main}, the main result of this paper.

\section{The Weighted Vertex-Independent Matching Lemma}\label{sec:vertexindependentlemma}
In this section, we turn to prove Lemma~\ref{lemma:vertex-independent} which was used in Section~\ref{sec:analysis}. We restate the lemma below and for simplicity of notation,  drop the primes in symbols such as $G', \mc{G}', \Delta'$ as stated in Section~\ref{sec:analysis} and use $G, \mc{G}, \Delta$ instead. 

\restate{Lemma~\ref{lemma:vertex-independent}.}{Let $G=(V, E, w)$ be an edge-weighted base graph with maximum degree $\Delta$. Let $\mc{G}$ be a random subgraph of $G$ that includes each edge $e \in E$ independently with some fixed probability $p \in (0, 1]$. Let $\mc{A}(H)$ be any (possibly randomized) algorithm that given any subgraph $H$ of $G$, returns a (not necessarily maximum weight) matching of $H$. For any $\epsilon > 0$ there is a randomized algorithm $\mc{B}$ to construct a matching $\inm{} = \mc{B}(\mc{G})$ of $\mc{G}$ such that
\begin{enumerate}
\item For any vertex $v$, $\Pr_{\mc{G}\sim G, \mc{B}}[v \in \inm{}] \leq \Pr_{\mc{G}\sim G, \mc{A}}[v \in \mc{A}(\mc{G})] + \epsilon^3.$
\item $\E[w(\inm{})] \geq (1-\epsilon)\E[w(\mc{A}(\mc{G}))]$ \item For any vertex-subset $\{v_1, v_2, \ldots \} \subseteq V$ such that for all $i, j$, $d_G(v_i, v_j) \geq \lambda$ where $\lambda = \O{\epsilon^{-24} \log \Delta \cdot \poly(\log\log \Delta)}$, events $\{v_1 \in \inm{}\}, \{v_2 \in \inm{}\}, \{v_3 \in \inm{}\}, \ldots$ are all independent with respect to both the randomizations used in algorithm $\mc{B}$ and in drawing $\mc{G}$. 
\end{enumerate}
}

\smparagraph{Outline of the proof.} To prove this lemma, we need to design an algorithm  $\mc{B}(\mc{G})$ that satisfies all three properties. If we only had the first two properties to satisfy, we could simply use algorithm $\mc{A}$. The problem however, becomes challenging when we need to, in addition, satisfy the third property regarding the independence   between  the events $\{v_1 \in \inm{}\}, \{v_2 \in \inm{}\}, \{v_3 \in \inm{}\}, \ldots$ for vertices $v_1, v_2, \ldots$, that are pair-wise far enough from each other. To ensure that our algorithm meets this condition, as it was done previously in the work of \cite{stoc20} for the unweighted variant of the lemma, we show that it can be implemented efficiently in the \local{} model of computation (whose formal description follows).

The \local{} model is a standard distributed computing model which consists of a network (graph) of processors with each processor having its own tape of random bits. Computation proceeds in synchronous rounds and in each round, processors can send unlimited size messages to each of their neighbors . Thus, to transmit a message from a node $u$ to node $v$, we require at least $d(u, v)$ rounds. For the same reason, if an algorithm terminates within $r$-rounds of \local{}, the output of any two nodes that have distance at least $2r$ from each other would be independent, which is essentially how we guarantee our independence property.

For simplicity, we explain our algorithm in a sequential setting in Algorithm~\ref{alg:YY}, and later describe how it can be simulated in the \local{} model.  We define a recursive algorithm $\findmatching{r}{\mc{G}}$ that given a parameter $r$, as the depth of recursion, and a subgraph of $G$, denoted by $\mc{G}$ outputs a matching of this graph. We give an informal overview of the algorithm in Section~\ref{section:overview}, and formally state it Section~\ref{section:algorithm}.

\smparagraph{Comparison to \cite{stoc20}.} 
For the proof, we follow the general recipe of \cite{stoc20} for the unweighted variant.  However, in this work we face several new challenges which make design and the analysis of the algorithm more complicated. Most importantly, the previous work relies on two fundamental observations which do not hold in this work. First, in unweighted graphs, if there exist two constant numbers $\delta$ and $\sigma$ such that for a $(1-\delta)$ fraction of the vertices $v\in V$ the following equation holds $$\Pr_{\mc{G}\sim G, \mc{B}} [v\in \mc{B}(\mc{G})]\geq (1-\sigma) \Pr_{\mc{G}\sim G, \mc{A}}[v \in \mc{A}(\mc{G})],$$ then we have $\E[|\mc{B}(\mc{G})|] \geq (1-\sigma)\E[|\mc{A}(\mc{G}))|] - \delta n$. Evidently, this only holds for the size of the matching but not for its weight. Second, as a result of the sparsification lemma in the previous work (which we discuss in Section~\ref{sec:techniques}), they could assume $|\mc{A}(\mc{G})| = \Omega(n)$.  Subsequently, to prove that $\mc{B}(\mc{G})$ provides a $(1-\epsilon)$-approximation, they only needed to show that $\sigma$ and $\delta$ are small enough constants.  As we discussed in Section~\ref{sec:techniques}, the sparsification lemma does not hold for weighted graphs. Thus, we need to take a completely different approach in our analysis.

 \subsection{Overview of the Algorithm}\label{section:overview}
We define a recursive algorithm $\findmatching{r}{\mc{G}}$ that given a parameter $r$, as the depth of recursion, and a subgraph of $G$, denoted by $\mc{G}$ outputs a matching of this graph. We then set our algorithm $\findmatching{}{\mc{G}} := \findmatching{t}{\mc{G}}$ for a number $t = O(\epsilon^{-20})$.
For $r=0$, algorithm $\findmatching{0}{\mc{G}}$ simply returns an empty matching. For any $r>0$, the idea is to use the matching constructed in $\findmatching{r-1}{\mc{G}}$ and transform it to a one that is sufficiently heavier in expectation. However, this transformation needs to be in a way that the probability of a vertex being matched in $\findmatching{r}{\mc{G}}$ is not significantly higher than $\Pr_{\mc{G}\sim G, \mc{A}}[v \in \mc{A}(\mc{G})]$.  A useful observation here is that we do not need to ensure that for any given subgraph $\mc{G}$ algorithm  $\findmatching{}{\mc{G}}$ gives a large enough matching while the probability of a vertex being matched in the algorithm is not greater than $\Pr_{\mc{G'}\sim G, \mc{A}}[v \in \mc{A}(\mc{G'})] + \epsilon^3$, rather we need this to hold in expectation over realization of $\mc{G}$. We strongly use this observation in the design of our algorithm by drawing several ($\epsilon^{-12}$) other random realization of $G$ and simultaneously constructing a matching for each one. This way, we have the freedom of matching a vertex with a high probability in an instance, in the expense of the vertex being matched with a lower probability in another instance. Similarly, we might construct a relatively low-weight matching for an instance but compensate it by finding a relatively heavier matching in another one. More precisely, in $\findmatching{}{\mc{G}}$, we have  $\alpha = \epsilon^{-12}+1$ random realizations of $G$, denoted by $\mc{G}_1, \dots, \mc{G}_\alpha$, where $\mc{G}_1= \mc{G}$, and our goal is to construct matchings $M'_1, \dots, M'_\alpha$ for them simultaneously. Roughly speaking, since our input subgraph $\mc{G}$ is itself a random realization of $G$ and that all these subgraphs are drawn from the same distribution, we  achieve our goal if our algorithm performs as desired in average over these $\alpha$ realizations. 

Below we provide a definition which we will use to refer to our subgraphs and their corresponding matching. 

\begin{definition}[profiles]\label{def:profile}
We say $((\mc{\cru{}}_1, M_1 ), \dots, (\mc{\cru{}}_k, M_k ))$ is a profile of size $k$, iff for any $i\in[k]$, $\mc{\cru{}}_i$ is a subgraph of $\cru{}$ and $M_i$ is a matching on $\mc{\cru{}}_i$.
\end{definition}

To construct matchings $M'_1, \dots, M'_\alpha$ for subgraphs  $\mc{G}_1, \dots, \mc{G}_\alpha$ in algorithm $\findmatching{r}{\mc{\mc{G}}}$, we start by running $\findmatching{r}{\mc{G}_i}$ for any $i\in [\alpha]$, and obtain matchings $M_1, \dots, M_\alpha$ as a result. In the other words, we start from profile $((\mc{\cru{}}_1, M_1 ), \dots, (\mc{\cru{}}_\alpha, M_\alpha))$ and want to transform it to $((\mc{\cru{}}_1, M'_1 ), \dots, (\mc{\cru{}}_\alpha, M'_\alpha))$ such that $\E[w(M'_i)]$ is sufficiently greater than $\E[w(M_i)]$ for a random $i\in [\alpha]$, while the constraints in the second and third properties of  Lemma~\ref{lemma:vertex-independent} are not violated. To get this, we use an idea similar to finding augmenting paths in the classic weighted matching algorithms. However, ours rather than being a path, is a structure that consists of multiple paths in graphs $\mc{G}_1, \dots, \mc{G}_\alpha$. We call this structure a \emph{multi-walk} and formally define it  in Definition~\ref{def:multi-walk}. Similar to how augmenting paths are used, we will use this structure to flip the membership of some edges in their corresponding matchings with the goal of increasing the expected size of the matchings.  However, note that if we naively choose the multi-walks with the sole purpose of increasing the average size of the matchings, we might violate the second property of lemma, as it might lead to some vertices being matched with an undesirably large probability. Further, these multi-walks should not include vertices that are further than a threshold since otherwise we might violate the third property of the lemma.
To overcome the first issue, after probability of a vertex $v$ being matched in our algorithm reaches a threshold, we mark it as \emph{saturated}. When a vertex is saturated, our algorithm ensures that while augmenting the matchings (using multi-walks), it does not increase the number of matchings in which this vertex is matched.   Having these constrains narrows down our choices of augmenting structures (multi-walks) significantly. However, we give a constructive proof (using Algorithm~\ref{alg:constructH}), and show that this narrow set includes a subset that can be used to increase the average size of our matchings sufficiently.  

\subsection{Algorithm $\mc{B}(\mc{G})$}\label{section:algorithm}
We start by providing some definitions that will be used in the Algorithm.
\begin{definition}[multi-walks]\label{def:multi-walk}
We define $W=((s_1, e_1), \dots, (s_l, e_l))$ to be a multi-walk of length $l$ of profile 
 $P=((\mc{\cru{}}_1, M_1 ), \dots, (\mc{\cru{}}_k, M_k ))$ iff it satisfies the following conditions.
  \begin{itemize}
 	\item For any $i\in [l]$, we have $s_i \in [k]$, and $e_i$ is an edge in subgraph $\mc{\cru{}}_{s_i}$.  
 	 \item  $(e_1, \dots, e_k)$ is a walk in graph $G$.
 	 \item W contains distinct elements, e.g., for any $i$ and $j$, we have $(s_i, e_i) \neq (s_j, e_j)$.  
 \end{itemize}
 Given a profile  $P=((\mc{\cru{}}_1, M_1 ), \dots, (\mc{\cru{}}_j, M_k ))$ and a multi-walk $W= ((s_1, e_1), \dots, (s_l, e_l))$,  we say $P \oplus W = ((\mc{\cru{}}_1, M'_1 ), \dots, (\mc{\cru{}}_j, M'_k ))$ is the result of applying $W$ on $P$ iff for any $i\in [k]$, $M'_i$ is constructed as follows:
$$M'_i = M_i \cup \{e_j \,|\, i = s_j \text{ and } e_j\notin M_i \} \backslash \{e_j \,|\,  i = s_j \text{ and } e_j\in M_i \}. $$ 
\end{definition}

\begin{definition}[alternating multi-walks]\label{def:alternating}
	A multi-walk $W=((s_1, e_1), \dots, (s_k, e_k))$ of profile $P=((\mc{\cru{}}_1, M_1 ), \dots, (\mc{\cru{}}_\alpha, M_\alpha ))$ is an alternating multi-walk iff it satisfies the two following conditions. First, for any $i\in [k-1]$ we have $\pmb{1}(e_i\in M_{s_i}) + \pmb{1}(e_{i+1}\in M_{s_{i+1}}) =1$, and second, $P \oplus W$ is a profile.  	
	We further define $g(W, P)$, the gain  of applying alternating multi-walk $W$ on $P$, as $$\gain(W, P) =  \sum_{(i, e')\in W} ( \pmb{1}(e'\notin M_i)- \pmb{1}(e'\in M_i))w(e').$$
\end{definition}
Given an alternating multi-walk  $W=((s_1, e_1), \dots, (s_l, e_l))$ of  $P= ((\mc{\cru{}}_1, M_1 ), \dots, (\mc{\cru{}}_k, M_k ))$, and any vertex $v\in V$ we define $d_{W,v}$ and $\bar{d}_{W,v}$ as follows: 
\begin{equation} d_{W,v} = |\{i: v\in e_i \text{, and } e_i\in M_{s_i} \}| \; \text{ and }\; \bar{d}_{W,v} = |\{i: v\in e_i  \text{, and } e_i \notin M_{s_i} \}|. \label{eq:jehuvfr}\end{equation}

\begin{definition}[applicable multi-walks]\label{def:kfo3rff}
Given a multi-walk $W=((s_1, e_1), \dots, (s_l, e_l))$ of  profile $P$, and a subset of vertices $V_s$, we say $W$ is \emph{applicable} with respect to a set of vertices $V_s$ iff it is alternating and for any $v\in V_s$ it satisfies $d_{W,v} \geq \bar{d}_{W,v}$.
\end{definition}

To prove Lemma~\ref{lemma:vertex-independent}, we design an algorithm $\mc{B}$ that given a random realization of $G$ outputs a matching $Z$ and show that it satisfies the desired properties of the lemma. In \ref{alg:YY}, we provide a recursive algorithm $\findmatching{r}{\mc{\cru{}}}$ that given an integer number $r$ and a realization $\mc{\cru{}}$ of $\cru{}$ outputs a matching of $\mc{\cru{}}$. We set  $\mc{B}(\mc{\cru{}}) = \findmatching{t}{\mc{\cru{}}}$ for $t=c_t\epsilon^{-20}$ where $c_t$ is a constant number. (We fix the value of $c_t$ later.)

\begin{tboxalg2e}{\findmatching{r}{\mc{\cru{}}}}
\begin{algorithm}[H]
	\DontPrintSemicolon
	\SetAlgoSkip{bigskip}
	\SetAlgoInsideSkip{}
	\label{alg:YY}
	If $r=0$,  return an empty matching.\;
	Set $\alpha \leftarrow \epsilon^{-12}+1$, $l \leftarrow 3\epsilon^{-3}$.\;
	For any $i\in[\alpha]$, construct $\mc{\cru{}}_i$ as follows. We set $\mc{\cru{}}_1 := \mc{\cru{}}$, and for any $ 1< r$ subgraph $\mc{\cru{}}_i$ includes any edge $e\in \cru{}$ independently with probability $p$. \label{line:3}\\;
	Define profile $P := ((\mc{\cru{}}_1, M_1), \dots,  (\mc{\cru{}}_\alpha, M_\alpha))$ where $M_i := \findmatching{r-1}{\mc{\cru{}}_i}$. \label{line:4}\;
	Call a vertex $v$ {\em saturated} iff	$\Pr_{\mc{G'}\sim G, \mc{B}}[v \in \inm{}_{r-1}] \leq \Pr_{\mc{G}\sim G, \mc{A}}[v \in \mc{A}(\mc{G'})] + \epsilon^3 - 1/\alpha,$ and  {\em unsaturated} otherwise. \\ \label{line:6}
	Let $\mathcal{W}_a$ be the set of alternating multi-walks of $P$ that are applicable with respect to the set of saturated vertices. \label{line:add1}\;
	Construct the weighted hyper-graph $H=(V, E_H)$ as follows. For any multi-walk $W$ in set $\mathcal{W}_a$ with length at most $l$, $H$ contains a hyper-edge between vertices in $W$ with weight $\gain{(W, P)}$. \label{line:setI}\;
	$M_H \gets \apxMM{H}$. \algcomment{See  Proposition~\ref{alg:harris} for the $\apxMM$ algorithm.}\label{line:alghariss}\;
	Iterate over all hyper-edges in $M_H$, apply their corresponding multi-walks on $P$, and let $P' := ((\mc{\cru{}}_1, M'_1), \dots,  (\mc{\cru{}}_\alpha, M'_\alpha))$ be the final profile. \label{line:1-8}\;
	Return matching $M'_1$.
\end{algorithm}
\end{tboxalg2e}

\begin{observation}\label{obs:samedis}
For any $r$, matchings $M'_1, \dots, M'_\alpha$ in Algorithm~$\findmatching{r}{\mc{G}}$ are random variables that are drawn from the same distribution.	
\end{observation}
\begin{proof}
This is due to the fact that matchings $M_1, \dots, M_\alpha$ are independent random variables from the same distribution, and that to obtain $M'_1, \dots, M'_\alpha$, based on these matchings, algorithm does not treat them differently.
\end{proof}

Before proceeding to the proof of the three properties let us prove the following lemma about alternating multi-walks.
\begin{lemma}\label{lemma:njkfrjf}
Let $W = ((s_1, e_1), \dots, (s_k, e_{k}))$  be a multi-walk
of profile $P=((\mc{\cru{}}_1, M_1 ), \dots, (\mc{\cru{}}_j, M_\alpha))$    with $e_i = (u_i, u_{i+1})$ for any $i\in [k]$. If $W$ is an alternating multi-walk, then it satisfies the following properties:
\begin{enumerate} 
\item For any $v\in V$, if $v\notin \{u_1, u_{k+1}\}$, then we have $d_{W,v} = \bar{d}_{W,v}$. \label{item:first}
\item If $e_1 \in M_{s_1}$, then we have $d_{W,u_1} \geq \bar{d}_{W,u_1}$. Also, if $e_1 \notin M^{}_{s_1}$, we have  $d_{W,u_1} \leq \bar{d}_{W,u_1}$. \label{item:second}
\item If $e_1 \in M_{s_1}$ and $e_k \in M_{s_k}$, then $W$ is applicable with respect to any subset of $V$.\label{item:third}
\end{enumerate}
\end{lemma}

\begin{proof} 
 Observe that for any $i>1$, we have $u_i\in e_i$ and $u_i\in e_{i-1}$.  
 Consider an arbitrary vertex $v\in V$. Since $W$ is alternating, for any $1 <j \leq k$ that $v = u_j$, we either have $e_{j-1} \in M_{s_{j-1}}$ and  $e_j \notin M_{s_j}$ or $e_{j-1} \notin M_{s_{j-1}}$ and $e_j \in M_{s_j}$.   This implies: 
  $$d_{W,u_1}  = 
 |\{i: 1<i \leq k, v = u_i\}| +  \pmb{1}(v=u_1, e_1\in M_{s_1})+  \pmb{1}(v=u_{k+1}, e_k\in M_{s_k}),$$ 
 and
  $$\bar{d}_{W,u_1}  = 
 |\{i: 1<i \leq k, v = u_i\}| +  \pmb{1}(v=u_1, e_1\notin M_{s_1})+  \pmb{1}(v=u_{k+1}, e_k\notin M_{s_k}).$$
 Note that if $v\notin \{u_1, u_{k+1}\}$, then we have $$ d_{W,v}= | \{i: v\in e_i  \text{, and } e_i \in M_{s_i} \}| = 
 |\{i: 1<i \leq k, v = u_i\}| = |\{i: v\in e_i  \text{, and } e_i \notin M_{s_i} \}|= \bar{d}_{W,v},$$ which completes the proof of the first item. To prove the second item, note that if $e_1 \in M_{s_1}$, then we have $\pmb{1}(u_1=u_1, e_1\in M_{s_1})= 1$ and $\pmb{1}(u_1=u_1, e_1\notin M_{s_1})= 0$, which gives us $$\pmb{1}(u_1=u_1, e_1\in M_{s_1})+  \pmb{1}(u_1=u_{k+1}, e_k\in M_{s_k}) \geq \pmb{1}(u_1=u_1, e_1\notin M_{s_1})+  \pmb{1}(u_1=u_{k+1}, e_k\notin M_{s_k}),$$ and results in $d_{W,u_1} \geq \bar{d}_{W,u_1}$.  A similar argument shows that if $e_1 \notin M_{s_1}$, then $d_{W,u_1} \leq \bar{d}_{W,u_1}$ holds.

 Since multi-walks are not directed the second claim of the lemma can also be interpreted as follows. If $e_k \in M_{s_k}$ then, $d_{W,u_{k+1}} \ge \bar{d}_{W,u_{k+1}}$.
 Combining this with the first claim of the lemma, we obtain that if $e_1 \in M_{s_1}$, and $e_k \in M_{s_k}$, then for any $v\in V$, we have $d_{W,v} \geq \bar{d}_{W,v}$. By definition of applicable multi-walks, this means that if $e_1 \in M_{s_1}$, and $e_k \in M_{s_k}$ then multi-walk $W$ is applicable with respect to any subset of $V$. This completes the proof the lemma.
 \end{proof}

\subsection{The First Property of Lemma \ref{lemma:vertex-independent}: Matching Probabilities}
In this section our goal is to prove that Algorithm $\mc{B}(\mc{G})$ satisfies the first property of Lemma~\ref{lemma:vertex-independent} as follows. 
\begin{lemma}
For any vertex $v\in V$, we have $\Pr_{\mc{G}\sim G, \mc{B}}[v \in \inm{}] \leq \Pr_{\mc{G}\sim G, \mc{A}}[v \in \mc{A}(\mc{G})] + \epsilon^3$.	
\end{lemma}
\begin{proof}
We will prove a stronger claim which is for any $v\in V,$ and any $r\leq t$, we have $q_{r, v} \leq q^{\mc{A}}_v + \epsilon^3,$ where 
$$q_{r, v} := \Pr_{\mc{G}\sim G, \mc{B}}[v \in \findmatching{r}{\mc{\cru{}}}], \;\; \text{ and } \;\; q^{\mc{A}}_v := \Pr_{\mc{G}\sim G, \mc{B}}[v \in \mc{A}(\mc{G})].$$
We use proof by induction. The claim obviously holds for $r=0.$ For any $r>0$, we assume that $q_{r-1, v} \leq q^{\mc{A}}_v + \epsilon^3$ holds and obtain $q_{r, v} \leq q^{\mc{A}}_v$. Draw a random realization of $G$ and denote it by $\mc{G}$ (i.e. $\mc{G}\sim G$). Consider matchings $M_i, \dots, M_{\alpha}$, and $M'_i, \dots, M'_{\alpha}$ from algorithm $\findmatching{r}{\mc{G}}$, and let us define
$$\rho_{r, v}:= |\{i: v\in M_i\}|/\alpha, \;\;\; \text{and} \;\;\; \rho'_{r, v} = |\{i: v\in M'_i\}|/\alpha.$$ 
We claim that $q_{r-1, v} = \rho_{r, v}$ and $q_{r, v} = \rho'_{r, v}$ hold. The former is due to the fact that any $i\in [\alpha]$, $M_i$ is the result of running algorithm $\mc{B}_{r-1}$ on a random realization of $G$ which by definition is equal to $q_{r-1, v}$. For the latter, note that we have $M'_i = \findmatching{r}{\mc{G}}$ and by Observation~\ref{obs:samedis}, we know that matchings $M'_i, \dots, M'_{\alpha}$ are drawn from the same distribution. As a result, we get $$|\{i: v\in M'_i\}| = \alpha \Pr[v\in \findmatching{r}{\mc{G}}],$$ which implies $q_{r, v} = \rho'_{r, v}$.

We prove our induction step for the cases of $q_{r-1,v} \leq q^{\mc{A}}_v + \epsilon^3 - 1/\alpha,$  and $q_{r-1,v} > q^{\mc{A}}_v + \epsilon^3 - 1/\alpha$ separately. We first show that if $q_{r-1,v} \leq q^{\mc{A}}_v + \epsilon^3 - 1/\alpha,$ (i.e., $v$ is not saturated), then $\rho_{r, v} \geq \rho'_{r,v}- 1/\alpha$ holds, which can be interpreted as $$q^{\mc{A}}_v + \epsilon^3 - 1/\alpha \geq q_{r-1, v} \geq q_{r, v} - 1/\alpha,$$ and as a result $q^{\mc{A}}_v + \epsilon^3 \geq q_{r, v}$.  Let $W_H$ denote the set of multi-walks corresponding to edges in $M_H$ constructed in  $\findmatching{r}{\mc{G}}$. Since $M_H$ is a matching, for any vertex $v$, there exists at most one multi-walk $W\in W_H$ that contains vertex $v$. In addition, since $W$ is alternating, we have $|d_{W, v} - \bar{d}_{W, v}| \leq 1,$ where $d_{W, v}$ and  $\bar{d}_{W, v}$ are defined as $$d_{W,v} = |\{i: v\in e_i \text{, and } e_i\in M_{s_i} \}| \; \text{ and }\; \bar{d}_{W,v} = |\{i: v\in e_i  \text{, and } e_i \notin M_{s_i} \}|.$$  
Since after applying a multi-walk $W$ on a profile, membership of the edges in $W$ flips in their corresponding matchings, we get $|\{i: v\in M_i\}| \geq |\{i: v\in M'_i\}|-1$ which means $\rho_{r, v} \geq \rho'_{r,v} - 1/\alpha$. We now consider the case of $q_{r-1,v} \geq  q^{\mc{A}}_v + \epsilon^3 - 1/\alpha,$ (i.e., $v$ is saturated) and show that in this case, $\rho_{r, v} \geq \rho'_{r,v}$ holds.  Due to $W$ being applicable with respect to the set of saturated vertices, by Definition~\ref{def:kfo3rff}, it satisfies $d_{W, v} \geq \bar{d}_{W, v}$. This directly yields  $\rho_{r, v} \geq \rho'_{r,v}$, and as a result $q_{r-1, v} \geq q_{r,v}$. Based on the induction hypothesis, we have $q_{r-1, v} \leq q^{\mc{A}}_v + \epsilon^3$ which implies   $q_{r,v}\leq q^{\mc{A}}_v + \epsilon^3$ and completes the proof.
\end{proof}

\subsection{The Second Property of Lemma~\ref{lemma:vertex-independent}: 
Expected Weight of the Matching}
In this section, our goal is to prove  $\E[w(\inm{})] \geq (1-\epsilon)$, where $\inm{} = \findmatching{t}{\mc{G}}$ for $t=c_t\epsilon^{-20}$. We will fix the value of the constant $c_t$ later in this section.

We start by Lemma~\ref{lemma:hahah} concerning the relation between the expected weight of the matching  and the weight of matching $M_H$ on hyper-graph $H$ in the algorithm. For any $r$, let $M_{H,r}$ denote the matching $M_H$ in algorithm $\findmatching{r}{\mc{G}}$.

\begin{lemma}\label{lemma:hahah}
 For any $0<r\leq t$, we have $\E_{\mc{G}\sim G}[\findmatching{r}{\mc{G}}] =\E_{\mc{G}\sim G} [\findmatching{r-1}{\mc{G}}] +\E[w(M_{H,r})]/\alpha.$
\end{lemma}
\begin{proof}
Consider algorithm $\findmatching{r}{\mc{G}}$ where $\mc{G}$ is a random realization of $G$. To prove this lemma, we will show 
\begin{equation}\sum_{i\in \alpha} w(M'_i) - \sum_{i\in \alpha} w(M_i)  = w(M_{H,r}). \label{eq:hheyhi}\end{equation}
By Algorithm~\ref{alg:YY}, we have $\findmatching{r}{\mc{G}} = M'_1$. Moreover, Observation~\ref{obs:samedis} states that matchings  $M'_1, \dots, M'_{\alpha}$ are all drawn from the same distribution which implies $$\E\Bigg[\sum_{i\in \alpha} w(M'_i)\Bigg] = \alpha \E_{\mc{G}\sim G}[\findmatching{r}{\mc{G}}]. $$
Similarly, since matchings $M_1, \dots, M_{\alpha}$ are all drawn from the same distribution as $\findmatching{r-1}{\mc{G}}$ we have  
$$\E\Bigg[\sum_{i\in \alpha} w(M_i)\Bigg] = \alpha \E_{\mc{G}\sim G}[\findmatching{r-1}{\mc{G}}]. $$
Consequently, to prove the lemma, it suffices to prove Equation~\ref{eq:hheyhi} holds. Let $W_H$ denote the set of multi-walks corresponding to edges in $M_H$ constructed in  $\findmatching{r}{\mc{G}}$. Since the weight of each edge in $H$ is equal to the gain of its corresponding multi-walk, we can write
\begin{equation}w(M_{H}) = \sum_{W\in W_H} g(W, P) = \sum_{W\in W_H} \sum_{(i, e)\in W} ( \pmb{1}(e\notin M_i)- \pmb{1}(e\in M_i))w(e).\label{eq:oiuefhiuwe}\end{equation}
Note that profile $P'$ is the result of iteratively applying the set of multi-walks $W_H$ on profile $P$. However, since $M_{H}$ is a matching, and as a result multi-walks in $W_H$ are vertex disjoint, gain of a multi-walk is not affected by the multi-walks applied before that. Moreover, since different multi-walks concern different vertices of the graph, we can assume w.l.o.g, that we apply all of them at the same time. Let us define for any $i\in [\alpha],$  
$$E_{i,1} =  \bigcup_{W\in W_H} \{e \,|\, (i, e)\in W \text{ and } e\notin M_i \}, \;\; \text{ and }\;\;
E_{i,2} =  \bigcup_{W\in W_H} \{e \,|\, (i, e)\in W \text{ and } e\in M_i \}. $$ 
By Definition~\ref{def:profile}, for any $i\in[\alpha]$, we have $M'_i = M_i \cup E_{i,1} \backslash E_{i,2}.$ This implies 

$$w(M'_i) - w(M_i)= \sum_{e\in E_{i,1}} w_e - \sum_{e\in E_{i,2}} w_e = \sum_{W\in W_H} \sum_{(j, e) \in W, j=i} ( \pmb{1}(e'\notin M_j)- \pmb{1}(e'\in M_j))w(e'),$$
and as a result
$$\sum_{i\in [\alpha]}w(M'_i) - w(M_i)= \sum_{W\in W_H} \sum_{(j, e) \in W} ( \pmb{1}(e'\notin M_j)- \pmb{1}(e'\in M_j))w(e').$$
Combining this with Equation~\ref{eq:oiuefhiuwe} results in Equation~\ref{eq:hheyhi} and completes the proof.
\end{proof}
For any $r\leq t$, let $Z_r :=\findmatching{r}{\mc{G}}$. Given Lemma~\ref{lemma:hahah}, to prove the second property, it suffices to show that for any $r$ having $\E[w(\inm{}_r)] < (1-\epsilon)\E[w(\mc{A}(\mc{G}))]$ results in $\E[w(M_{H,r})] \geq \alpha \E[w(\mc{A}(\mc{G}))]/t$. Based on Lemma~\ref{lemma:hahah}, this implies  $$\E[w(\inm{}_t)]\geq \sum_{r <t}\E[w(M_{H,r})]/\alpha  \geq  \min\left(t\alpha \E[w(\mc{A}(\mc{G}))]/(t\alpha), (1-\epsilon)\E[w(\mc{A}(\mc{G}))]\right) = (1-\epsilon)\E[w(\mc{A}(\mc{G})),$$ which is equivalent to the second property of Lemma~\ref{lemma:vertex-independent}.
To achieve this, in Lemma~\ref{lemma:lemlemuih} (stated below), we prove that having $\E[w(\inm{})] < (1-\epsilon)\E[w(\mc{A}(\mc{G}))]$ results in  $\E[w(M_{H,r})] = \Omega(\epsilon^8\E[w(\mc{A}(\mc{G}))]),$ 
which can be interpreted as $\E[w(M_{H,r})] \geq c\epsilon^8\E[w(\mc{A}(\mc{G}))]$ for a constant number $c$. By setting $$c_t = \frac{\epsilon^{-12}+1}{c\epsilon^{-12}},$$ we get 
$$\E[w(M_{H,r})] = c\epsilon^8\E[w(\mc{A}(\mc{G}))] = \frac{(\epsilon^{-12}+1)\E[w(\mc{A}(\mc{G}))]}{c_t\epsilon^{-20}}.$$
Recall that we have $t=c_t\epsilon^{-20}$, and $\alpha=\epsilon^{-12}+1$, which gives us $\E[w(M_{H,r})] \geq \alpha\E[w(\mc{A}(\mc{G}))]/t$. Therefore, to prove the second property of Lemma~\ref{lemma:vertex-independent}, it only suffices to prove the following lemma. 
\begin{lemma}\label{lemma:lemlemuih}
For any $r\leq t$, if $\E[w(\inm{})] < (1-\epsilon)\E[w(\mc{A}(\mc{G}))]$, then  $\E[w(M_{H,r})] = \Omega(\epsilon^8\E[w(\mc{A}(\mc{G}))]).$ 
\end{lemma}
\begin{proof}
To prove this, we will construct a subgraph $H'$ of $H$ which max-degree $2\alpha$ such that
$$
\E\left[\sum_{e\in H'} w(e)\right] \geq \alpha \epsilon^2 \E[w(\mc{A}(\mc{G}))].
$$ First, note that $H$ is a hyper-graph of rank $l=3\epsilon^{-3}$ since each edge is between the vertices of a path of length at most $l$ in $G$.
Using Lemma~\ref{lemma:hupergraphsize}, we know that subgraph $H'$ (and as a result hyper-graph $H$) has a matching of weight $\sum_{e\in H'} w(e)/(2l\alpha)$ which is in expectation equal to $\epsilon^5 \E[w(\mc{A}(\mc{G}))]/6.$
 Moreover, $M_{H,r}$ is constructed by $\apxMM{H}$ which by Proposition~\ref{alg:harris} returns an $O(l)$-approximation of the maximum weight matching of $H$. Thus, we get $$\E[w(M_{H,r})] = \Omega(\epsilon^8\E[w(\mc{A}(\mc{G}))]) .$$ 

Before proceeding to the construction of $H'$ in Algorithm~\ref{alg:constructH}, let us provide some definitions. Given a profile $P=((\mc{\cru{}}_1, M_1 ), \dots, (\mc{\cru{}}_j, M_k ))$, we say  $W = ((s_1, e_1), \dots, (s_a, e_a))$, an alternating multi-walk of $P$, is \emph{expandable} by $W' =((s'_1, e'_1), \dots, (s'_b, e'_b))$ iff either $W_1$ or $W_2$, defined below, is an alternating multi-walk:
$$W_1 = ((s_1, e_1), \dots, (s_a, e_a), (s'_1, e'_1), \dots, (s'_b, e'_b)),$$$$W_2=((s'_1, e'_1), \dots, (s'_b, e'_b), (s_1, e_1), \dots, (s_a, e_a)).$$
If $W$ is expandable by $W'$ either one of $W_1$ and $W_2$ that is an alternating multi-walk is the result of expanding $W$ by $W'$. (If both are alternating multi-walks, we pick one arbitrarily.) Similarly, we say $W$ is expandable by a path or a cycle $p = (e'_1, \dots, e'_b)$  in graph $G_i$ iff $W$ is expandable by $((i, e'_1), \dots, (i, e'_b))$, and the result of expanding $W$ by $p$ is similar to expanding $W$ by $((i, e'_1), \dots, (i, e'_b))$. 

Below we state Algorithm~\ref{alg:constructH} which given profile $P$ and the set of saturated vertices $V_s$ outputs hyper-graph $H'$. Note that both $P$ and $V_s$ are from algorithm \findmatching{r}{\mc{G}} by which $M_{H,r}$ is constructed.

\begin{tboxalg2e}{Constructing subgraph $H'$ given profile $P := ((\mc{\cru{}}_1, M_1), \dots,  (\mc{\cru{}}_\alpha, M_\alpha))$ and $V_s$.}
\begin{algorithm}[H]
	\DontPrintSemicolon
	\SetAlgoSkip{bigskip}
	\SetAlgoInsideSkip{}
	\label{alg:constructH}
	Define $H'$ to be a hyper-graph with vertex set $V$ that initially does not have any  edges.\\
For any $i\in [\alpha]$, let $M^{\mc{A}}_i := \mc{A}(\mc{G}_i)$, and
 $E'_i := \{e\in \mc{G}_i \,|\, \pmb{1}(e\in M_i) + \pmb{1}(e\in M^{\mc{A}}_i)=1 \}$  \algcomment{$E'_i$  contains an edge if it is in exactly one of $M_i$ and $M^{\mc{A}}_i$.}
\\
Let 	$V_r := \{v\in V_s :  |\{i: v\in M^{\mc{A}}_i  \}| > |\{i: v\in M_i  \}|\}$. \\
	
Remove an edge $e$ from $E'_i$ iff $e\in M^{\mc{A}}_i$ and at least one of its end-points is in  $V_r$. \\ 	\label{line:5}
Let $\mc{G}'_i := (V, E'_i)$. \label{line:6}\\
	\While{$\text{there exists an } i\in {\alpha}, \text{where }  E'_i \neq \emptyset$, \label{line:10}}{
		Let $W$ be an empty multi-walk.\\
		Pick a maximal path or a cycle $p$ from $\mc{G}'_i$. \label{line:7}\\
		If $W$ is expandable by $p$, expand $W$ by $p$, and and remove all the edges of $p$ from $E'_j$. \label{line:8}\\
		\While{there exists a subgraph $\mc{G}'_j$ that contains a maximal path or a cycle $p$ by which $W$ is expandable,}{
			Expand $W$ by $p$ and remove all the edges of $p$ from $E'_j$. \label{line:12}
	}
	Add $W$ to $\mc{W}$.
	}
	\For{any $W \in \mc{W}$,} 
	{Pick an integer number $x$ between $0$ and $l/4-1$ uniformly at random. \\
	Decompose $W = ((s_1, e_1), \dots, (s_k, e_k))$ to smaller multi-walks $W_1, \dots, W_a$ by removing any element $(s_i, e_i)$ from the multi-walk iff $e_i \notin M_{s_i}$ and either $i\bmod (l/4) = x$ or $i\bmod (l/4) = x+1$ hold. \label{line:16} \\  
	If $W_1$ is expandable by $W_a$, expand $W_1$ by $W_a$, and set $W_a$ to be an empty multi-walk.\\
	For any multi-walk $W' \in \{W_1, \dots W_a\}$, add an edge to hyper-graph $H'$ between the vertices in $W'$ with weight $\gain(W')$. \label{line:17}\\
	}
		Return $H'$.	
\end{algorithm}
\end{tboxalg2e}
To complete the proof of Lemma~\ref{lemma:lemlemuih}, we need to show that hyper-graph $H'$ outputted by Algorithm~\ref{alg:constructH}, has the three following properties.
\begin{enumerate}
\item The maximum degree of hyper-graph $H'$ is upper-bounded by $2\alpha$.
\item hyper-graph $H'$ is a subgraph of hyper-graph $H$.
\item We have $\E[\sum_{e\in H'} w(e)] \geq \alpha\epsilon^2\E[w(\mc{A}(\mc{G}))]$. 
\end{enumerate}
For the first property of $H'$ first observe that any hyper-edge $e\in H'$  represents a multi-walk $W_e$ in $P$. For any vertex $v$, if $v\in e$, then $W_e$ contains an element $(i, e')$ where $v\in e'$ and $e'\in \mc{G}'_i$.  Moreover, in the algorithm, after using $(i, e')$ in construction of a multi-walk, we remove $e'$ from subgraph $\mc{G}'_i$. (see Line~\ref{line:12} of Algorithm~\ref{alg:constructH}.) We also know that degree of each vertex in $\mc{G}'_i$ is at most two. This gives us an upper-bound of $2\alpha$ for degree of each vertex in $H'$.

To prove the second property, let us first recall that based on Line~\ref{line:setI} of Algorithm~\ref{alg:YY}, hyper-graph $H$ has a hyper-edge for any multi-walk of length at most $l$ in set $\mc{W}_a$ (which is defined as the set of multi-walks of $P$ that are applicable with respect to the set of saturated vertices). To prove this property, it suffices to show that any hyper-edge in $H'$ also represent a multi-walk of length at most $l$ in $\mc{W}_a$. Since in both graphs $H$ and $H'$, weight of each edge is set to be the gain of its corresponding multi-walk, we do not need to consider the edge-weights in our proof. Consider a multi-walk $W'$ from Line~\ref{line:17} of Algorithm~\ref{alg:constructH}. Since any edge in $H'$ represents a multi-walk described in this line of the algorithm, to complete the proof we only need to show that $W'$ is a multi-walk of lenght at most $l$ in $\mc{W}_a$. Clearly, the length of this multi-walk is at most $l$ due to Line~\ref{line:16} of Algorithm~\ref{alg:constructH}.  Moreover, Lemma~\ref{lemma:jihihi} states that $W'$ is an alternating multi-walk and is applicable with respect to the saturated vertices, which implies $W'\in \mc{W}_a$,  and completes the proof of this property.

To give a lower-bound for $\E[\sum_{e\in H'} w(e)]$ we will prove that $$\E\left[\sum_{e\in H'} w(e)\right] \geq \alpha((1-3\epsilon^3)\E[w(\mc{A}(\mc{G}))] - \E[w(Z)]),$$ which considering  $\E[w(\inm{})] < (1-\epsilon)\E[w(\mc{A}(\mc{G}))]$ in the statement of lemma results in:

$$\E\left[\sum_{e\in H'} w(e)\right] \geq \alpha(\epsilon-3\epsilon^3)\E[w(\mc{A}(\mc{G}))].$$ For a small enough $\epsilon$ that satisfies $\epsilon^2 > \epsilon-3\epsilon^3$ we can write this as  
$$\E\left[\sum_{e\in H'} w(e)\right] \geq \alpha\epsilon^2\E[w(\mc{A}(\mc{G}))],$$ which is equivalent to the third property of $H'$.
For any $e\in H'$ ,let $W_e$ be the multi-walk in Line~\ref{line:17} of Algorithm~\ref{alg:constructH} represented by $e$.
By definition of $g(W_e, P)$, and the fact that for any $(i, e')\in W_e$, if  $e'\notin M_i$, then $e'\in M^{\mc{A}}_i$ we get:
 $$w(e)= g(W_e, P) = \sum_{(i, e')\in W_p} ( \pmb{1}(e'\notin M_i)- \pmb{1}(e'\in M_i))w(e') = \sum_{(i, e')\in W_p} ( \pmb{1}(e'\in M^{\mc{A}}_i)- \pmb{1}(e'\in M_i))w(e').$$
 Observe that based on $Algorithm~\ref{alg:constructH}$, for any $i\in[\alpha]$ and any edge $e'\in M_i$, there exists an edge $e\in H'$ such that $(i, e')\in W_e$. Similarly, for any $i\in[\alpha]$ and any edge $e'\in M^{\mc{A}}_i$, there exists an edge $e\in H'$ such that $(i, e')\in W_e$ unless $e'$ is removed in Line~\ref{line:5} of the algorithm or $(i, e')$ is removed in Line~\ref{line:16} of the algorithm. Based on Lemma~\ref{lemma:kjfnkef} we know that probability of $e'$ being removed in Line~\ref{line:5} is upper-bounded by $\epsilon^3$. Moreover, it is easy to see that probability of $(i, e')$ being removed in Line~\ref{line:16} is upper-bounded by $4/l$ = $4\epsilon^3/3$. This means that with probability of at least $1-3\epsilon^3$, for any $i\in[\alpha]$ and any edge $e'\in M^{\mc{A}}_i$, there exists an edge $e\in H'$ such that $(i, e')\in W_e$. This implies
 $$\E\left[\sum_{e\in H'} w(e)\right] = \sum_{i\in \alpha} \left( \sum_{e'\in M^{\mc{A}}_i} (1-3\epsilon^3) w(e') - \sum_{e'\in M_i} w(e')\right) = \sum_{i\in \alpha} \left( (1-3\epsilon^3) w(M^{\mc{A}}_i) - w(M_i)\right).$$
Since matchings $M_1, \dots, M_\alpha$ are drawn from the same distribution, and similarly, matchings $M^{\mc{A}}_1, \dots, M^{\mc{A}}_\alpha$ are drawn from the same distribution, for any $i\in [\alpha]$ we have  
  $\E[w(M_i)] = \E[w(Z_r)]$ and $\E[w(M^{\mc{A}}_i)] = \E[w(\mc{A}(\mc{G}))]$. This gives us  $$\E\left[\sum_{e\in H'} w(e)\right] = \alpha\left( (1-3\epsilon^3) \E\left[w(\mc{A}(\mc{G}))\right] - \E[w(Z_r)]\right),$$
and concludes the proof of this Lemma.  
\end{proof}

\begin{lemma}\label{claim:iu3rh}
Consider multi-walks $\{W_1, \dots, W_a\}$  in Line~\ref{line:17} of Algorithm~\ref{alg:constructH}. If there exists an $i\in [a]$, where $W_i$ is not applicable with respect to set $V_s$, then $W$ is not applicable with respect to this set either. 
\end{lemma}
\begin{proof}
 We use proof by contradiction. We assume that $W = ((s_1, e_1), \dots, (s_k, e_{k}))$ is an alternating multi-walk applicable with respect to set $V_s$ while there exists an $i\in[a]$ where $W_i$ is not applicable with respect to this set. We then show that this leads to a contradiction. If $W_i$ is not applicable with respect to $V_s$, then either it is not alternating, or there exists a vertex $v \in V_s$ for which $d_{W_i, v} < \bar{d}_{W_i, v}$. By Lemma~\ref{lemma:njkfrjf}, if $W$ is alternating then any $v\in V$ that satisfies $d_{W_i, v} < \bar{d}_{W_i, v}$ is an end-point of $W_i$. Therefore, to obtain a contradiction, it suffices to prove that $W$ is alternating, and that if $v\in V_s$ is an end-point of $W$, then  $d_{W_i, v} \geq \bar{d}_{W_i, v}$. 

We first prove our claim for the case of $1<i<a$. By construction, in this case, $W_i$ is a subsequence of $W$, i.e.,  $W_i = ((s_x, e_x), \dots, (s_y, e_y))$ for $1 < x < y < k$, and as a result it is an alternating multi-walk. We will show that in this case, multi-walk $W$ is applicable with respect to any subset of $V$. Based on Lemma~\ref{lemma:njkfrjf}, to get this, it suffices to show that $e_x \in M_{s_x}$ and $e_y\in M_{s_y}$ hold. Since $W_i$ is a result of decomposing $W$, we know that elements $(s_{x-1}, e_{x-1})$ and $(s_{y+1}, e_{y+1})$ are removed in Line~\ref{line:16} of the algorithm. As a result we have $e_{x-1}\notin M_{s_{x-1}}$ and $e_{y-1}\notin M_{s_{y-1}}$. Combining this with the fact that $W$ in alternating, we get $e_{x} \in M_{s_{x-1}}$ and $e_{y} \in M_{s_{y}}$.

To complete the proof, it remains to show that for any $i\in \{1, a\}$, multi-walk $W_i$ is alternating, and that any vertex $v$ which is an end-point of $W_i$ satisfies $d_{W_i, v} \geq \bar{d}_{W_i, v}$. For any $i\in [k]$, let $e_i = (u_i, u_{i+1})$ which means that for any $i>1$, we have $u_i\in e_{i-1}$ and $u_i\in e_i$. Consider the multi-walks $W_1$ and $W_a$ in  Line~\ref{line:16} of the algorithm. We assume w.l.o.g. that during the decomposing of $W$ to shorter multi-walks, it is decomposed to at least two multi-walks and as a result $1<a$. At this point of the algorithm,  we have $W_1 = ((s_1, e_1), \dots, (s_x, e_x))$ and $W_a = ((s_y, e_y), \dots, (s_k, e_k))$ for some $1 \leq x <  y \leq k$. Note that both $W_1$  and $W_k$ are alternating  multi-walks due to being subsequences of $W$. Moreover, similar to the previous case, we can argue that $e_x \in M_{s_x}$ and $e_y \in M_{s_y}$ due to the fact that elements $(s_{x+1}, e_{x+1})$ and $(s_{y+1}, e_{y+1})$ are removed during the decomposition process. If we also have $e_1 \in M_{s_x}$ and $e_k \in M_{s_x}$ then $W_1$ is not expandable by $W_a$ and both these multi-walks are applicable with respect to any set of vertices due to the third item of  Lemma~\ref{lemma:njkfrjf}. Therefore, we focus on the case that either $e_k \notin M_{s_x}$ or $e_1 \notin M_{s_x}$ holds. Let us assume w.l.o.g. that we have $e_k\notin M_{s_x}$. It is easy to see that if $u_1\notin V_s$ then $W_1$ is applicable with respect to $V_s$.
We claim that in this case of $e_k\notin M_{s_x}$, if $u_1\in V_s$, then we have $u_1 = u_{k+1}$ and $e_1\in M_{s_x}$ as otherwise $W$ does meet the condition  $d_{W_i, v} \geq \bar{d}_{W_i, v}$ which is necessary for $W$ being applicable with respect to set $V_s$. This implies that $W_1$ is expandable by $W_a$ since $((s_1, e_1), \dots, (s_x, e_x), (s_y, e_y), \dots, (s_k, e_k))$ is an alternating multi-walk. As a result to complete the proof we only need to show that the result of expanding  $W_1$ by $W_a$ is applicable with respect to $V_s$. Indeed in this case, this multi-walk is applicable with respect to any set of vertices due to $e_x \in M_{s_x}$ and $e_y \in M_{s_y}$ and the third item of Lemma~\ref{lemma:njkfrjf}. Thus, the proof of the this lemma is concluded.
\end{proof}

\begin{lemma}\label{claim:oi3rui}
The while loop in Line~\ref{line:10} of Algorithm~\ref{alg:constructH} terminates and $\mc{W}$ constructed by that is a set of alternating multi-walks. \end{lemma}
\begin{proof}

It is easy to see that if the loop terminates $\mc{W}$ only contains alternating multi-walks since any multi-walk $W$ added to this set is the result of iteratively expanding an empty multi-walk by a set of paths and cycles. Recall that by definition, an empty multi-walk is alternating and the result of expanding an alternating multi-walk by a path or a cycle is also an alternating multi-walk. The while loop terminates when for any $i\in [\alpha]$, we have  $E'_i = \emptyset$, thus to complete the proof, it suffices to show that each iteration of the loop terminates and that in each one, we remove at least one edge from one of the subgraphs $\mc{G}'_1, \dots, \mc{G}'_\alpha$. We consider an arbitrary iteration of the loop, and show that in Line~\ref{line:8}, edges of $p$ are removed from $\mc{G}'_i$. This happens iff $W$ is expandable by $p$. Multi-walk $W$ is empty at this point of the algorithm (and as a result is an alternating multi-walk) and $p = (e_1, \dots, e_k)$ is a maximal (nonempty) path or a cycle chosen from an arbitrary $\mc{G}'_i$ in Line~\ref{line:7}. As an application of Lemma~\ref{lemma:oiej}, we get that $W$ is expandable by $p$. As a result of this, in Line~\ref{line:8} of the algorithm edges of $p$ are removed from $E'_i$. To conclude that the while loop terminates we also have to show that each of its iterations terminate. It is easy to see since the loop nesting in this while loop obviously terminates as well.
\end{proof}

\begin{lemma}\label{lemma:oiej} Let $p =(e'_1, \dots,  e'_b)$ be a a maximal connected-component (a path or a cycle) in graph $\mc{\cru{}'}_i$ (defined in Algorithm~\ref{alg:constructH}), and 
let $W = ((s_1, e_1), \dots, (s_a, e_a))$ be an alternating multi-walk of profile $P' = ((\mc{\cru{}}_1, M_1), \dots,  (\mc{\cru{}}_\alpha, M_\alpha))$, such that for any $j\in [b]$, we have $(i, e'_j)\notin W$ and for any $j\in [a]$, we have $e_j\in E'_{s_j}$. If the first vertex of $W$ is the same as the last vertex of $p$ and $\pmb{1}(e_1\in M_{s_1}) + \pmb{1}(e'_b\in M_{i}) =1$, then $W$ is expandable by $p$.
\end{lemma}

\begin{proof}
 First, let us note that any maximal connected-component in graph $\mc{\cru{}'}_i$ is a path or a cycle since we have $E'_i\subset (M_i \cup M^{\mc{A}}_i)$, and as a result the degree of each vertex in $\mc{G}'_i$ is at most two. (Recall that, $M_i$ and $M^{\mc{A}}_i$ are both matchings of graph $\mc{G}_i$.)
To prove that $W$ is expandable by $p$ we will show that $W_p=((i, e'_1), \dots,  (i, e'_b), (s_1, e_1), \dots, (s_a, e_a))$ is an alternating multi-walk. First, $W_p$ is a multi-walk since $(e'_1, \dots,  e'_b, e_1, \dots, e_a)$  is a walk in $G$ and it also contains distinct elements as for any $j\in [b]$, $(i, e'_j)\notin W$ holds. 

By Definition~\ref{def:alternating}, to prove that $W_p$ is alternating, we first need to show that for any two consecutive elements in $W_p$, e.g., $(s''_1, e''_1)$ and $(s''_2, e''_2)$, we have $\pmb{1}(e''_1\in M_{s''_1}) + \pmb{1}(e''_2\in M_{s''_2}) =1$. If both these elements are in $W$ this simply holds due to $W$ being an alternating multi-walk itself. Moreover, if exactly one of them is in $W$, we get this as a result of $\pmb{1}(e_1\in M_{s_1}) + \pmb{1}(e'_b\in M_{i}) =1$ (in the statement of lemma). Therefore, we need to focus on showing that for any $j\in [b-1]$, we have $\pmb{1}(e_j\in M_{i}) + \pmb{1}(e_{j+1}\in M_{i}) =1$. Since $E'_i\subset (M_i \cup M^{\mc{A}}_i)$ and by the fact that  $M_i$ and  $M^{\mc{A}}_i$ are matchings of graph $\mc{G}_i$, if $e_i\in M_i$ then  $e_{i+1}\notin M_i$. Similarly, if $e_i\notin M_i$ then $e_{i}\in M^{\mc{A}}_i$ which gives us $e_{i+1}\notin M^{\mc{A}}_i$ and  $e_{i+1}\in M_i$. 
 
As the second condition in Definition~\ref{def:alternating}, we need to show that $P \Delta W_p = ((\mc{\cru{}}_1, M'_1 ), \dots, (\mc{\cru{}}_k, M'_k ))$ is a profile, where for any $j\in [\alpha]$ we have 
\begin{equation} \label{eq:kuft} M'_j = M_j \cup \{e \,|\, (j, e)\in W_p   \text{ and } e \notin M_j \} \backslash \{e \,|\, (j, e)\in W_p  \text{ and } e\in M_j \}.\end{equation} 
By Definition~\ref{def:profile}, to prove that $P \Delta W_p$ is a profile, it only suffices to show that for any $j\in [\alpha]$, $M'_j$ is a matching in $\mc{G}_j$. This simply holds for any $j\neq i$ due to $W$ being an alternating multi-walk itself, thus we only need to show that $M_i$ is a matching in $\mc{G}_i$. To achieve this, we consider any two edges $\{e, e'\} \subset M'_i$ and show that $e$ and $e'$ are not adjacent in $\mc{G}_i$. If neither one of these edges is in $p$, then for $W$ to be an alternating multi-walk these edges cannot be adjacent. Moreover, it is easy to see that if both edges are in $p$, they are not adjacent either. Thus, we assume that exactly one of the edges is in $p$. W.l.o.g., we assume $e\in p$ and $e'\notin p$. We consider two cases of $e'\in G'_i$ and $e'\notin G'_i$. In the first case, $e$ and $e'$ are not adjacent since $p$ is a maximal component of $G'_i$ and as a result is not connected to edges that are not in $p$ (including $e'$). In the case of $e'\notin G'_i$, we claim that $e'$ is in both $M_i$ and $M^{\mc{A}}_i$ which means it cannot be adjacent to any edge in $G'_i$ including $e$. To prove this claim, note that by Equation~\ref{eq:kuft}, we have $M'_i \subset (M_i \cup \{e'' \,|\, (i, e'')\in W_p\})$ and by the statement of lemma
for any $(i, e'')\in W_p$ we have $e''\in E'_i$. Moreover, by definition of $\mc{G}'_i$, we know $E'_i \subset (M_i \cup M^{\mc{A}}_i)$. Putting these facts together results in the following equation:
$$M'_i \subset (M_i \cup \{e'' \,|\, (i, e'')\in W_p\}) \subset (M_i\cup E'_{i}) \subset (M_i \cup M^{\mc{A}}_i).$$ 
Recall that $\mc{G}'_i$ contains an edge iff it is in $(M_s \cup M^{\mc{A}}_a)$ but not in $(M_s \cap M^{\mc{A}}_a)$. As a result since e is in $M'_i$ but it is not in $\mc{G}'_i$, then it is in $(M_s \cap M^{\mc{A}}_a)$. This completes the proof of our lemma since we obtained that $W_p$ is an alternating multi-walk.
\end{proof}

\begin{claim}\label{claim:claimclaim}
In Line~\ref{line:6} of Agorithm~\ref{alg:constructH}, for any $v\in V_s$, we have  $r_v \geq g_v$ where
$ g_v = |\{i: v\in (M^{\mc{A}}_i \cap E_i')\}|$ and  $r_v = |\{i: v\in (M_i \cap E_i') \}|$.
\end{claim}

\begin{proof}
We use proof by contradiction. Let $v\in V_s$ be a vertex with $r_v < g_v$. It is easy to see that we have $v \notin V_r$ since in Line~\ref{line:5}, for any $i\in [\alpha]$, we remove any edge in $E'_i$  which has at least one end-point in $V_r$. As a result, in Line~\ref{line:6}, for any $u\in V_r$ we have $d_{v,g} = 0$. Due to $v \notin V_r$, we get $|\{i: v\in M^{\mc{A}}_i  \}| \leq |\{i: v\in M_i  \}|$.   Observe that for any  $v \notin V_r$, we have  $$|\{i: v\in (M^{\mc{A}}_i \cap E_i')\}| = |\{i: v\in M^{\mc{A}}_i\}| - |\{i: v\in (M^{\mc{A}}_i \cap M_i )\}| \text{, and }$$ $$|\{i: v\in (M_i \cap E_i')\}| = |\{i: v\in M_i\}| - |\{i: v\in (M^{\mc{A}}_i \cap M_i )\}|.$$ This gives us $ r_v -g_v=|\{i: v\in M_i \}|-|\{i: v\in M^{\mc{A}}_i \}|,$ which implies $r_v \geq g_v$ and completes our proof.
\end{proof}

%

\begin{lemma} \label{lemma:jihihi}
Any multi-walk in line~\ref{line:17} of Algorithm~\ref{alg:constructH} which is represented by an edge in hyper-graph $H'$ is applicable with respect to the vertices in $V_s$.
\end{lemma}

\begin{proof}
By Lemma~\ref{claim:iu3rh}, to prove this, it suffices to show that any $W \in \mc{W}$ constructed in the algorithm is applicable with respect to $V_s$. Recall that, by Definition~\ref{def:kfo3rff}, a multi-walk $W$ of profile $P$ is applicable with respect to $V_s$ iff it is alternating and it satisfies $d_{W,v} \geq  \bar{d}_{W,v}$ for any $v \in V_s$.  Based on Lemma~\ref{claim:oi3rui}, W is an alternating multi-walk thus it remains to show that for any $v\in V_s$, we have $d_{W,v} \geq  \bar{d}_{W,v}$.

We use proof by contradiction. We start by assuming that there exists a vertex $v\in V_s$ and a multi-walk $W'\in \mc{W}$ where $d_{W',v} < \bar{d}_{W',v}$ and then show that it results in a contradiction. Let $W = ((s_1, e_1), \dots, (s_k, e_{k}))$ be the first multi-walk for which we have $d_{W,v} \neq \bar{d}_{W,v}$. By Lemma~\ref{lemma:njkfrjf}, this implies that vertex $v$ is an endpoint of this multi-walk. W.l.o.g., let us assume that we have $e_1= (v, u_2)$. Consider subgraphs $\mc{G'}_1, \dots, \mc{G'}_{\alpha}$ in the algorithm when $W$ is added to $\mc{W}$. Due to the condition of the while loop in Line~\ref{line:10} of the algorithm the following holds at this point of the algorithm.  There does not exist a $\mc{G'}_i$ that contains a maximal path $p$ with which $W$ is expandable. By Lemma~\ref{lemma:oiej}, this implies that any maximal path $p = (e'_1, \dots, e'_a)$ in any subgraph $\mc{G'}_i$ that ends in vertex $v$ (i.e., $e'_a = (u'_a, v)$) satisfies $\pmb{1}(e_1\in M_{s_1}) = \pmb{1}(e'_a\in M_{i})$. We consider both  cases of  $e_1\in M_{s_1}$ and $e_1\in M_{s_1}$ and prove prove the lemma for each one independently.

Let us assume that $e_1\in M_{s_1}$. In this case, by Item~\ref{item:second} of Lemma~\ref{lemma:njkfrjf}, we have $d_{W, v} \geq \bar{d}_{W,v}$ which means $W\neq W'$. We will show that in this case, any multi-walk $W''$ added to set $\mc{W}$ in the next iterations satisfies $d_{W'', v} \geq \bar{d}_{W'',v}$ which contradicts the existence of $W'$. Consider a maximal connected component (a path or a cycle) $p = (e'_1, \dots, e'_a)$ in  $\mc{G}_i$ for an arbitrary $i\in [\alpha]$, and define $W_p = ((i, e'_1), \dots, (i,e'_a))$. By Lemma~\ref{lemma:oiej} $W_p$ is an alternating multi-walk. Moreover, by Item~\ref{item:first} of Lemma~\ref{lemma:njkfrjf} if $v$ is not an end-point of $p$ (which also includes the case that $p$ is a cycle) then we have $d_{W_, v} = \bar{d}_{W_p,v}$. Further, if $p$ is a path and $v$ is one of its end-points, i.e., $e'_a = (u'_a, v)$, as mentioned above we have $\pmb{1}(e_1\in M_{s_1}) = \pmb{1}(e'_a\in M_{i})$, which means $e'_a\in M_{i}$. As a result of this and by invoking the second item of Lemma~\ref{lemma:njkfrjf}, we get that $d_{W_p, v} \geq \bar{d}_{W_p,v}$. Note that any multi-walk $W''$ constructed in the next iterations consists of a set of maximal connected components. Since all the remaining connected components satisfy $d_{W_p, v} \geq \bar{d}_{W_p,v}$, we also have  $d_{W'', v} \geq \bar{d}_{W'', v}$. This contradicts the existence of multi-walk $W'$ with $d_{W', v} < \bar{d}_{W', v}$.

Now we consider the case of $e_1\notin M_{s_1}$. We will show that this assumption results in equation $ |\{i: v\in (M^{\mc{A}}_i \cap E_i')\}| < |\{i: v\in (M_i \cap E_i') \}|$ for vertex $v$, which contradicts the statement of Claim~\ref{claim:claimclaim}. First, we show that if $e_1\notin M_{s_1}$ then any multi-walk $W''\in \mc{W}$ satisfies $d_{W'', v} \leq \bar{d}_{W'',v}$. Let us consider a path or cycle $p = (e'_1, \dots, e'_a)$ in graph $\mc{G}_i$ for an arbitrary $i\in [\alpha]$, and define $W_p = ((i, e'_1), \dots, (i,e'_a))$. Similar to what we used in the proof of the previous case, if  $v$ is not an end-point of $W_p$ (which also includes the case of $p$ being a cycle), then by Lemma~\ref{lemma:njkfrjf}, we have $d_{W_p, v} \geq \bar{d}_{W_p,v}$. Moreover, if $p$ is a path and $v$ is an end-point in this path, i.e., $e'_a = (u'_a, v)$, we have $\pmb{1}(e_1\in M_{s_1}) = \pmb{1}(e'_a\in M_{i})$. Since in this case we have $e_1\notin M_{s_1}$, we get $e'_a\notin M_{i}$. As a result of this, Item~\ref{item:second} in Lemma~\ref{lemma:njkfrjf} gives us  $d_{W_p, v} \leq \bar{d}_{W_p,v}$. Based on an argument that we used for the previous case, this implies that any mutli-walk $W''$ that we add to $\mc{W}$ in the next iterations satisfies $d_{W'', v} \geq \bar{d}_{W'', v}$. Moreover, due to the  assumption that $W$ is the first multi-walk that for any $W''$ that is added to this set before $W$ we have $d_{W'', v} = \bar{d}_{W'', v}$. We also have  $d_{W, v} < \bar{d}_{W, v}$ as a result of assumption $e_1\notin M_{s_1}$ and the second item of Lemma~\ref{lemma:njkfrjf}. This gives us the following equation:
 \begin{equation}\sum_{W\in \mc{W}} (\bar{d}_{W, v} - d_{W, v}) = \sum_{W\in \mc{W}} (|\{(i, e)\in W: v\in e, e\notin M_{i} \}| - |\{ (i, e): v\in e, e\in M_{i} \}|) > 0.\label{eq:janjan}\end{equation}
 where the first equality is due to the definition of $\bar{d}_{W, v}$ and $d_{W, v}$.
Further, based on Lemma~\ref{claim:oi3rui}, we know that  
the while loop in Line~\ref{line:10} of Algorithm~\ref{alg:constructH} terminates. When this loop terminates, there is no $j\in [\alpha]$ where $\mc{G}'_j$ contains at least one edge. This means that for any $e\in E'_j$ element $(e, i)$ is in exactly one of the multi-walks in $\mc{W}$. Also, note that by construction, $E'_j\subset (M^{\mc{A}}_j \cup M_j)$. As a results we get the following equations for vertex $v$:
$$|\{i: v\in (M_i \cap E_i')\}| = \sum_{W\in \mc{W}} |\{ (i,e)\in W: v\in e , e\in M_i\}|, \text{ and}$$
$$|\{i: v\in (M^{\mc{A}}_i \cap E_i')\}| = \sum_{W\in \mc{W}} |\{ (i,e)\in W: v\in e , e\notin M_i\}|.$$
Combining this with Equation~\ref{eq:janjan}, we get: $$|\{i: v\in (M^{\mc{A}}_i \cap E_i')\}| - |\{i: v\in (M_i \cap E_i')\}|>0$$
which is in contradiction with the following equation by Claim~\ref{claim:claimclaim} for any $v\in V_s$:
$$ |\{i: v\in (M^{\mc{A}}_i \cap E_i')\}| \leq  |\{i: v\in (M_i \cap E_i') \}|.$$ \end{proof}

\begin{lemma} \label{lemma:hupergraphsize}
Any weighted hyper-graph $K = (G, E_K)$ of max-degree $\Delta$ and rank $r$ has a matching with weight at least $\frac{1}{r\Delta}\sum_{e\in E_K} w(e)$.
\end{lemma}
\begin{proof}
We construct a matching $M_K$ using an iterative greedy algorithm and show that its weight is at least $\frac{1}{2\Delta}\sum_{e\in E_K} w(e)$. At the beginning all the edges are alive. In each iteration, we add an edge $e$ to $M_K$ which has the maximum weight among the alive edges and kill all its neighboring edges (that are not already killed by another vertex). Note that each edge $e$ in $M_K$ kills at most $r\Delta-1$ other edges with weight smaller than $w(e)$, which means $\sum_{e\in M_K} w(e) \geq \frac{1}{r\Delta} \sum_{e\in E_K} w(e)$.

\end{proof}

\begin{lemma}\label{lemma:kjfnkef}
Given that an edge $e = (u_1, u_2)$ exists in $M^{\mc{A}}_i$ defined in Algorithm~\ref{alg:constructH}, probability of this edge being removed in  Line~\ref{line:5} of the algorithm is upper-bounded by $\epsilon^3$. 
\end{lemma}
\begin{proof}
	Note that $e = (u_1, u_2)$ is removed in Line~\ref{line:5} of the algorithm iff $e\in M^{\mc{A}}_i$ and there exists a vertex $v\in \{u_1, u_2\}$ which is saturated and satisfies $|\{j: v\in M^{\mc{A}}_j  \}| < |\{j: v\in M_j  \}|$.  Let $I_e$ be an indicator random variable for the event of $e$ being removed from $\mc{G}'_i$ in Line~\ref{line:5} of the algorithm. Moreover, let us define $g_v:=|\{j: v\in M^{\mc{A}}_j  \}|$ and $r_v:= |\{j: v\in M_j  \}|$. We have 
		\begin{align}\Pr[I_e] \leq &\Pr[g_{u_1} > r_{u_1} \,|  \, u_1 \in  M^{\mc{A}}_i] + \Pr[g_{u_2} > r_{u_2} \,|\, u_2\in M^{\mc{A}}_i]. \label{eq:oi3ri3}
	 \end{align} 
Thus, it suffices to show that, $\Pr[g_v > r_v \,|\, v \in  M^{\mc{A}}_i]\leq \epsilon^3/2$ holds for any vertex $v \in \{u_1, u_2\}$. We have \begin{align} \Pr[g_v > r_v |\,v \in  M^{\mc{A}}_i]\leq \Pr[g_{v, -i} +1  > r_{v, -i} ]\leq \Pr[g_{v, -i} \geq  r_{v, -i}] \label{eq:ijefj}\end{align}
where $g_{v, -i}:=|\{j: j\neq i \text{ and } v\in M^{\mc{A}}_j  \}|$ and $r_{v,-i}:= |\{j: j\neq i \text{ and } v\in M_j  \}|$.
Recall that by definition of saturated vertices in Line~\ref{line:6} of Algorithm~\ref{alg:YY}, for any saturated vertex $v$ and $i\in [\alpha]$, we have $  \Pr[ v \in M_i]-  \Pr[ v \in M^{\mc{A}}_i] \geq \epsilon^3-1/\alpha$ and as a result $\E[r_{v, -i}] - \E[g_{v, -i}] \geq  (\alpha -1)(\epsilon^3-1/\alpha)$. To complete the proof, we show $$\Pr[|g_{v, -i} - \E[g_{v, -i}]|> (\alpha-1) \epsilon^4] \leq \epsilon^{-4}, \text{ and }\Pr[|r_v - \E[r_v]|> (\alpha-1) \epsilon^4]\leq \epsilon^{-4}.$$
Note that $g_{v, -i}$ and $r_{v, -i}$ are both sum of independent Bernoulli random variables as for any  $a$ and $b$, $\mc{G}'_a$ and $\mc{G}'_b$ are independent random variables. Therefore, to bound $\Pr[|g_{v, -i} - \E[g_{v, -i}]|> (\alpha-1) \epsilon^4]$ and $\Pr[|r_{v,-i} - \E[r_{v, -i}]|> (\alpha-1) \epsilon^4]$ we can use Chebyshev's inequality which states for any $k$, $\Pr[|r_{v,-i} - \E[r_{v, -i}]|>  \var(r_v)^{1/2}k] \leq k^{-2}$. Observe that $\var(r_{v,-i}) < (\alpha-1)$ and $\var(g_{v, -i}) < (\alpha-1)$. Based on Algorithm~\ref{alg:YY}, we have $\alpha-1 = \epsilon^{-12}$. This implies that 
\begin{align}\Pr[|r_{v,-i} - \E[r_{v, -i}]|>  (\alpha-1) \epsilon^4] \nonumber  & = \Pr[|r_{v,-i} - \E[r_{v, -i}]|>  \epsilon^{-8}] \nonumber   \\ \nonumber & =\Pr[|r_{v,-i} - \E[r_{v, -i}]|>  (\alpha-1)^{1/2} \epsilon^{-2}] ] \\ & \leq \nonumber \Pr[|r_{v,-i} - \E[r_{v, -i}]| \geq \var{(r_v)}^{-1/2} \epsilon^{-2}] ] \\ & \leq \epsilon^4. \nonumber \end{align}
We can similarly show that $\Pr[|g_{v,-i} - \E[g_{v, -i}]|>  (\alpha-1) \epsilon^4] \leq \epsilon^4$. Moreover, since  $\E[r_{v, -i}] - \E[g_{v, -i}] \geq  (\alpha -1)(\epsilon^3-1/\alpha)$, if $g_{v, -i} \geq  r_{v, -i}$ then, we either have $g_{v, -i} \geq \E[g_{v, -i}]+ (\alpha-1)(\epsilon^3-1/\alpha)/2$ or $r_{v, -i} \leq \E[r_{v, -i}]-(\alpha-1)(\epsilon^3-1/\alpha)/2.$ For a small enough $\epsilon$, we have $(\epsilon^3-1/\alpha)/2 \geq \epsilon^4$, and $$\Pr[g_{v, -i} \geq  r_{v, -i}] \leq \Pr[|r_{v,-i} - \E[r_{v, -i}]|>  (\alpha-1) \epsilon^4] + \Pr[|g_{v,-i} - \E[g_{v, -i}]|>  (\alpha-1) \epsilon^4]\leq 2\epsilon^4.$$ 
Combining this with Equation~\ref{eq:ijefj} and Equation~\ref{eq:oi3ri3} results in $\Pr[I_e] \leq 4\epsilon^4$ which for a small enough $\epsilon$, gives us $\Pr[I_e] \leq \epsilon^3$.   
\end{proof}

\subsection{The Third Property of Lemma \ref{lemma:vertex-independent}: Independence}
In this section our goal is to prove the following lemma.
\begin{lemma}
For any $0 \leq r\leq t$, algorithm $\findmatching{r}{\mc{\cru{}}}$ can be simulated in $O(\epsilon^{-24}\log\Delta \poly(\log\log\Delta))$ rounds of \local{}. 
\end{lemma}
\begin{proof}
We will show that for any $r\leq t$, algorithm $\findmatching{r}{\mc{\cru{}}}$ can be implemented in 	$$x_r := cr\epsilon^{-4}\log\Delta \poly(\log\log\Delta)$$ rounds of \local{} for a large enough constant $c$. Since we have $t=c_t\epsilon^{-20}$ for a constant $c_t$, this implies that $\findmatching{}{\mc{\cru{}}} = \findmatching{t}{\mc{\cru{}}}$ can be simulated in $O(\epsilon^{-24}\log\Delta \poly(\log\log\Delta))$  rounds. To prove this claim, we use proof by induction. As the base case, $\findmatching{0}{\mc{\cru{}}}$ can be simply implemented in $O(1)$ rounds as it only returns an empty matching. As the induction step, for any $r >1$, we assume that our claim holds for  $\findmatching{r-1}{\mc{\cru{}}}$, and prove that it holds for $\findmatching{r}{\mc{\cru{}}}$ too.

 Graph $G$ is the underlying graph in our \local{} simulation of $\findmatching{r}{\mc{\cru{}}}$,  and there is a processor on each $v\in V$. The initial information that each node $v$ holds is as follows.  Its incident neighbors in graphs $G $ and $\mc{G}$,  $\Pr_{\mc{G}\sim G, \mc{A}}[v \in \mc{A}(\mc{G})]$, and parameters $\epsilon$, $r$ and $\Delta$ (maximum degree of $G$).   Observe that other than $\mc{G}$, the rest of the initial information is independent of the realization of $\mc{G}$ and the randomization of the algorithm. Thus, if two vertices are not adjacent in $G$, they initially do not share any information that is correlated with the randomization of the algorithm or the realization of $\mc{G}$. As a result, to prove our lemma, we only need to show that using this initialization, we can implement our algorithm in the desired number of rounds. To prove our claim, we go over Algorithm~\ref{alg:YY} line by line, and investigate the number of rounds that we need to simulate each one in the \local{} model. The first two lines obviously take $O(1)$ round since no communication is needed for initializing the variables.
 
 In Line~\ref{line:3} and Line~\ref{line:4} of the algorithm, the goal is to construct profile $P$. First, to construct subgraphs $\mc{G}_2,\dots, \mc{G}_\alpha$, for any edge $e\in G$, we only need its end-points to communicate and hold the information about realization of $e$ in these subgraphs. This can be done in $O(1)$. Moreover, by the induction step for any $i$, algorithm $\findmatching{r-1}{\mc{G_i}}$ can be simulated in $x_{r-1}$ rounds. Further, $\findmatching{r-1}{\mc{G}_1}\dots, \findmatching{r-1}{\mc{G_{\alpha}}}$, can be constructed in parallel. As a result this line of the algorithm takes $x_{r-1} + O(1)$ rounds.
 
 To simulate Line~\ref{line:6} of the algorithm,  we show that any vertex  $v$ can compute $\Pr_{\mc{G'}\sim G, \mc{A}}[v \in \findmatching{r-1}{\mc{G'}}]$ and determine whether it is saturated or not after $x_{r-1}$ rounds of the algorithm. First, note that $\Pr_{\mc{G'}\sim G, \mc{A}}[v \in \findmatching{r-1}{\mc{G'}}]$ is just a function of $G$. Moreover, by the induction step, $\findmatching{r-1}{\mc{G'}}$ can be implemented in $x_{r-1}$ rounds of \local{}, which implies that $\Pr_{\mc{G'}\sim G, \mc{A}}[v \in \findmatching{r-1}{\mc{G'}}]$ is a function of $x_{r-1}$-hop of vertex $v$ in graph $G$. This is a piece of information that vertex $v$ can gather in $x_{r-1}$ rounds. Therefore, considering that initially each vertex holds the value of $\Pr_{\mc{G}\sim G, \mc{A}}[v \in \mc{A}(\mc{G})]$ and $\epsilon$, vertex $v$ can  determine whether it is saturated or not by evaluating the following inequality. 
  $$\Pr_{\mc{G}\sim G, \mc{B}}[v \in \inm{}_{r-1}] \leq \Pr_{\mc{G}\sim G, \mc{A}}[v \in \mc{A}(\mc{G})] + \epsilon^3 - 1/\alpha.$$
   This only adds an extra $O(1)$ to the round complexity of the  \local{} algorithm since each vertex can gather the necessary information during the $x_{r-1}+ O(1)$ that our algorithm has already run from the beginning of the algorithm. 
  
In Line~\ref{line:add1} and Line~\ref{line:setI}, the goal is to construct the hyper-graph $H$, which has a hyper-edge between the vertices of any multi-walk of length at most $l = 3\epsilon^{-3}$ of $P$ in set $\mathcal{W}_a$. Recall that $\mathcal{W}_a$ is the set of alternating multi-walks of $P$ that are applicable with respect to the set of saturated vertices. To achieve this, first, each vertex gathers all the information about the vertices in its $l$-hop and finds the alternating multi-walks of length at most $l$ that contain this vertex. In this way, each vertex knows all the edges of $H$ to which it belongs. This can obviously be done in $O(l)$ rounds. 

Line \ref{line:alghariss} of the algorithm is about  $\apxMM{H}$ which as mentioned before uses an algorithm by Harris~\cite{DBLP:conf/focs/Harris19} provided bellow.
\begin{proposition}[{\cite[Theorem~1.2]{DBLP:conf/focs/Harris19}}]\label{alg:harris} Given a hyper-graph of rank $r$ and a constant $\delta \in (0, 1/2)$, there is an $\tilde{O}(\log{\Delta} + r)$-round algorithm in the \local{} model to get an $O(r)$-approximation to maximum weight matching with probability at least $1-1/\delta$. Here the $\tilde{O}$ notation hides $\poly\log\log{\Delta}$ and $\poly\log{r}$ factors.
\end{proposition}
Based on this proposition, to analyze the round complexity of $\apxMM{H}$, we first  need to give an upper-bound for the maximum degree of $H$ which is the maximum number of hyper-edges in $H$ that any single vertex $v$ can belong to. In hyper-graph $H$, we have a hyper-edge between the vertices of any alternating hyper-walk $w=((s_1, e_1), \dots, (s_k, e_k))$ of length at most $l$ in profile $P$. By definition of multi-walks, $p = (e_1, \dots, e_k)$ should be a walk in graph $G$.  In a graph of maximum degree $\Delta$, there are at most $l\Delta^{l}$ distinct walks of length at most $l$ that contain vertex $v$. Further, for any $i\in [k]$, we have $s_i\in [\alpha]$ which means that there are at most $\alpha$ possible choices for any $s_i$. Thus, in graph $H$, there are at most $l(\Delta\alpha)^l$ edges that contain any arbitrary vertex $v$, and as a result maximum degree of $H$ is upper-bounded by $l(\Delta\alpha)^l$. Moreover, rank of hyper-graph $H$ is simply upper-bounded by $l$ since the rank of a hyper-graph is the maximum number of vertices that any edge contains. In the case of graph $H$ this is bounded by $l$ since each edge is between vertices of a walk of length at most $l$. 
Putting these together, and plugging in the value of variables $l=3\epsilon^{-3}$ and $\alpha=\epsilon^{-12}+1$, we obtain the following upper-bound for the round complexity of $\apxMM{H}$: $$\tilde{O}(\log{(\Delta\alpha)^{2l}} + l) = O(l\log{(\Delta\alpha)}\polylog{(l)}\poly\log\log{(l(\Delta\alpha)^{l}})) = O(\epsilon^{-4}\log{(\Delta)}\poly\log\log{(\Delta)}).$$ We can set the constant $c$ in a way that the number of rounds needed here is upper-bounded by $c\epsilon^{-4}\log{(\Delta)}\poly\log\log{(\Delta)}/2.$

Finally, in Line~\ref{line:12} we need to apply a set of multi-walks of length at most $l$ (constructed in previous rounds) on profile $P$. This can be easily done in $O(\epsilon^{-3})$-rounds since we have $l = 3\epsilon^{-3}$. To sum up, The overall round complexity of the algorithm which we denote by $R_r$ is as follows:
\begin{align}  R_r = & O(1) + x_{r-1} + O(l) + O(1) + O(l) + c\epsilon^{-4}\log{(\Delta)}\poly\log\log{(\Delta)}/2 \nonumber \\ = &  c(r-1)\epsilon^{-4}\log\Delta \poly(\log\log\Delta) + O(\epsilon^{-4}) + c\epsilon^{-4}\log\Delta \poly(\log\log\Delta)  \nonumber \\ =  & x_r \nonumber + O(\epsilon^{-4}) - c\epsilon^{-4}\log\Delta \poly(\log\log\Delta)/2.
\end{align}
Let $c_0\epsilon^{-4}$ be an upper-bound for what we denote in our round complexity as $O(\epsilon^{-4})$ where $c_0$ is constant.
We can set the constant $c$ to be large enough to satisfy 
$$c_0\epsilon^{-4} - c\epsilon^{-4}\log\Delta \poly(\log\log\Delta)/2 \leq 0. $$
This gives us $R_r \leq  x_{r}$, and concludes our proof.
\end{proof}
%
%
%
%

\bibliographystyle{plain}
\bibliography{refs}

\begin{thebibliography}{10}

\bibitem{sosa19}
Sepehr Assadi and Aaron Bernstein.
\newblock {Towards a Unified Theory of Sparsification for Matching Problems}.
\newblock In {\em 2nd Symposium on Simplicity in Algorithms, SOSA@SODA 2019,
  January 8-9, 2019 - San Diego, CA, {USA}}, pages 11:1--11:20, 2019.

\bibitem{AKL16}
Sepehr Assadi, Sanjeev Khanna, and Yang Li.
\newblock {The Stochastic Matching Problem with (Very) Few Queries}.
\newblock In {\em Proceedings of the 2016 {ACM} Conference on Economics and
  Computation, {EC} '16, Maastricht, The Netherlands, July 24-28, 2016}, pages
  43--60, 2016.

\bibitem{AKL17}
Sepehr Assadi, Sanjeev Khanna, and Yang Li.
\newblock {The Stochastic Matching Problem: Beating Half with a Non-Adaptive
  Algorithm}.
\newblock In {\em Proceedings of the 2017 {ACM} Conference on Economics and
  Computation, {EC} '17, Cambridge, MA, USA, June 26-30, 2017}, pages 99--116,
  2017.

\bibitem{sagt19}
Soheil Behnezhad, Mahsa Derakhshan, Alireza Farhadi, MohammadTaghi Hajiaghayi,
  and Nima Reyhani.
\newblock {Stochastic Matching on Uniformly Sparse Graphs}.
\newblock In {\em Algorithmic Game Theory - 12th International Symposium,
  {SAGT} 2019, Athens, Greece, September 30 - October 3, 2019, Proceedings},
  pages 357--373, 2019.

\bibitem{stoc20}
Soheil Behnezhad, Mahsa Derakhshan, and MohammadTaghi Hajiaghayi.
\newblock {Stochastic Matching with Few Queries: {$(1-\epsilon)$}
  Approximation}.
\newblock In {\em Proceedings of the 52nd Annual {ACM} {SIGACT} Symposium on
  Theory of Computing, {STOC} 2020, to appear}, 2020.

\bibitem{soda19}
Soheil Behnezhad, Alireza Farhadi, MohammadTaghi Hajiaghayi, and Nima Reyhani.
\newblock {Stochastic Matching with Few Queries: New Algorithms and Tools}.
\newblock In {\em Proceedings of the Thirtieth Annual {ACM-SIAM} Symposium on
  Discrete Algorithms, {SODA} 2019, San Diego, California, USA, January 6-9,
  2019}, pages 2855--2874, 2019.

\bibitem{BR18}
Soheil Behnezhad and Nima Reyhani.
\newblock {Almost Optimal Stochastic Weighted Matching with Few Queries}.
\newblock In {\em Proceedings of the 2018 {ACM} Conference on Economics and
  Computation, Ithaca, NY, USA, June 18-22, 2018}, pages 235--249, 2018.

\bibitem{blumetal}
Avrim Blum, John~P. Dickerson, Nika Haghtalab, Ariel~D. Procaccia, Tuomas
  Sandholm, and Ankit Sharma.
\newblock {Ignorance is Almost Bliss: Near-Optimal Stochastic Matching With Few
  Queries}.
\newblock In {\em Proceedings of the Sixteenth {ACM} Conference on Economics
  and Computation, {EC} '15, Portland, OR, USA, June 15-19, 2015}, pages
  325--342, 2015.

\bibitem{blumetalOR}
Avrim Blum, John~P. Dickerson, Nika Haghtalab, Ariel~D. Procaccia, Tuomas
  Sandholm, and Ankit Sharma.
\newblock {Ignorance Is Almost Bliss: Near-Optimal Stochastic Matching with Few
  Queries}.
\newblock {\em Operations Research}, 68(1):16--34, 2020.

\bibitem{edmonds1965maximum}
Jack Edmonds.
\newblock {Maximum matching and a polyhedron with 0, 1-vertices}.
\newblock {\em Journal of research of the National Bureau of Standards B},
  69(125-130):55--56, 1965.

\bibitem{DBLP:conf/focs/Harris19}
David~G. Harris.
\newblock Distributed local approximation algorithms for maximum matching in
  graphs and hypergraphs.
\newblock In David Zuckerman, editor, {\em 60th {IEEE} Annual Symposium on
  Foundations of Computer Science, {FOCS} 2019, Baltimore, Maryland, USA,
  November 9-12, 2019}, pages 700--724. {IEEE} Computer Society, 2019.

\bibitem{YM19}
Takanori Maehara and Yutaro Yamaguchi.
\newblock {Stochastic Monotone Submodular Maximization with Queries}.
\newblock {\em CoRR}, abs/1907.04083, 2019.

\bibitem{schrijver2003combinatorial}
Alexander Schrijver.
\newblock {\em {Combinatorial Optimization: Polyhedra and Efficiency}},
  volume~24.
\newblock Springer Science \& Business Media, 2003.

\bibitem{YM18}
Yutaro Yamaguchi and Takanori Maehara.
\newblock {Stochastic Packing Integer Programs with Few Queries}.
\newblock In {\em Proceedings of the Twenty-Ninth Annual {ACM-SIAM} Symposium
  on Discrete Algorithms, {SODA} 2018, New Orleans, LA, USA, January 7-10,
  2018}, pages 293--310, 2018.

\end{thebibliography}
	
\end{document}